\documentclass[aip,jmp,a4paper,
    amsmath,amssymb,
    reprint,onecolumn
    ]{revtex4-2}

\usepackage[utf8]{inputenc}
\usepackage[T1]{fontenc}
\usepackage{mathptmx}
\usepackage[colorlinks,hyperindex]{hyperref}
\hypersetup{
   colorlinks,
   citecolor=blue,
   linkcolor=blue,
   urlcolor=blue,
}
\usepackage{orcidlink}

\usepackage{amsthm}
    \theoremstyle{plain}

        \newtheorem{property}{Property}
        \newtheorem{corollary}{Corollary}

    \theoremstyle{definition}
        \newtheorem{definition}{Definition}

    \theoremstyle{remark}
        \newtheorem*{remark}{Remark}

\renewcommand{\geq}{\geqslant}

\newcommand{\fo}{\ensuremath{\mathfrak{o}}}

\DeclareMathOperator{\grad}{\operatorname{grad}}
\let\div\relax
\DeclareMathOperator{\div}{div}
\DeclareMathOperator{\curl}{curl}
\DeclareMathOperator{\STG}{STG}
\DeclareMathOperator{\STt}{\dot{\times}}
\newcommand{\N}{\nabla}

\begin{document}

\title{
    The conformally invariant symmetric traceless field:
    a conformal field strength and
    a conformally invariant gauge fixing equation 
    over (A)dS$_4$
}

\author{J.~Queva \orcidlink{0000-0001-8280-1925}}
\affiliation{Universit\'e de Corse -- CNRS UMR 6134 SPE, Campus Grimaldi BP 52, 20250 Corte, France.}
\email{queva@univ-corse.fr}

\begin{abstract}
    The conformally invariant symmetric traceless field $A$ is considered.
    In four dimensions it possesses a scalar gauge invariance to which
    we provide a conformally invariant gauge fixing equation.
    A field strength $F$ is built upon $A$, its 
    properties are worked out, giving rise to a set of 
    conformally invariant
    equations exhibiting an electromagnetic duality.
\end{abstract}

\maketitle

\section{Introduction}

Conformal symmetry is a recurring theme in physics, offering essential tools for both theoretical and mathematical physics. This symmetry is embodied in the source-free Maxwell's equations and within the standard model of particle physics, prior to the consideration of spontaneous symmetry breaking and quantum corrections. Additionally, it can emerge as an approximate symmetry in scenarios like high-energy processes in particle physics or near a critical point. Furthermore, conformal invariance may appear as an ``hidden'' symmetry in certain non-relativistic models, such as the hydrogen atom, and plays a crucial role in achieving a detailed comprehension of the spectrum within these models.

Conformal and gauge symmetries frequently arise together.
In many instances, the gauge symmetry is fixed while spoiling the conformal invariance.
However, this is not always the case; it is possible for conformal invariance to be preserved even when the gauge symmetry is fixed.
This article investigates such a case.

The results derived in the present paper are diverse, originating from the conformally invariant symmetric traceless field $A^{\mu_1\dots\mu_s}$ of rank $s$
fulfilling the conformally invariant equation on 
conformally flat Einstein spacetimes (CFES)
\begin{multline}\label{EqPrinc}
    (E_s(A))^{\mu_1\dots\mu_s}
    = (\square + c_sR) A^{\mu_1\dots\mu_s}
    + a_s s \nabla^{(\mu_1} \nabla_{\rho} A^{\mu_2\dots\mu_s)\rho} \\
    + b_s s(s-1)  g^{(\mu_1\mu_2} \nabla_{\rho}\nabla_{\sigma} A^{\mu_3\dots\mu_s)\rho\sigma}
    = 0,
\end{multline}
with $(\mu_1\dots\mu_s)$ is the normalized symmetrized part of the enclosed indices,
$\square = \nabla^\alpha\nabla_\alpha$ the laplacian, 
$R$ the scalar curvature of $(M,g)$ of dimension $d\geqslant 3$,
and the coefficients are
\begin{equation}
    \label{Coefficients}
    a_s = - \frac{4}{d + 2s-2},\quad
    b_s = \frac{4}{(d + 2s-4)(d + 2s-2)},\quad
    c_s = - \frac{d^2 -2d +4s}{4d(d-1)}.
\end{equation}
Fields such as $A$ could be of interest in the case of massless conformally
invariant higher spin fields and to generalizations of conformal gravity, and
in peculiar to such fields on (anti-)de Sitter spaces [(A)dS].

This article is organized as follows:
in Sec.~\ref{SecGuidTour} we examine fields of rank $1$ and $2$, exploring gauge freedom in dimension $d=4$  and the underlying structure between equations of different ranks on CFES.
In Sec.~\ref{SecPropEqMain} we confirm that conformal and gauge invariance for $d=4$ persist for fields of rank $s\geqslant 2$.
In Sec.~\ref{SecGaugFix} we demonstrate the conformal invariance of a gauge condition
which generalizes the Eastwood-Singer gauge fixing equation on CFES. 
To be precise, for $d=4$ and setting $\phi = \nabla_{\mu_1}\dots\nabla_{\mu_s} A^{\mu_1\dots\mu_s}$, we show the restricted conformal invariance of the set
\begin{equation}
    \label{Set1}
	\left\{\begin{aligned}
		& E_s(A) =0, \\
		& E_0(\phi ) = 0.
		\end{aligned}
		\right.
\end{equation}
In Sec.~\ref{SecFS} we investigate the properties of the field strength
$F=\mathcal{D}A$ associated to the field $A$.
Its gauge invariance is \emph{explicitly} shown and we establish that for $d=4$ 
it fulfills its own set of conformally invariant equations.
In Sec.~\ref{sec:E-B_decomposition}, the field strength $F$ is decomposed into $E$-$B$ fields with electromagnetic duality.
Using this duality, we demonstrate how the conformally invariant equations satisfied by $F$ translate into a pair of two complex conjugated conformally invariant equations on the $E$-$B$ fields
and how $\mathfrak{o}(2,4)$ acts on those fields.
In Sec.~\ref{sec:Conclusion}, an action giving
rise to Eq.~\eqref{EqPrinc} is retro-engineered on the field strength $F$ and a 
conformally invariant and gauge-invariant symplectic form is provided.
In App.~\ref{SecRemConf} we recall elementary formulas related to conformal invariance.
In App.~\ref{GFEqEff} we discuss the residual gauge freedom left by the conformally invariant gauge fixing equation.
In App.~\ref{Sec2D} we examine the lack of content of Eq.~\eqref{EqPrinc} in
the case $d=2$.
Finally, App.~\ref{sec:Proofs} compiles two lengthy proofs.

\section{Study of rank 1 and rank 2 fields}
\label{SecGuidTour}

The study of conformally invariant traceless fields of rank 1 and 2 reveals the properties of Eq.~\eqref{EqPrinc} and hints at what is to be expected for an arbitrary rank
$s\geq 2$ on CFES.
Moreover, it exposits in their simplest form the manner in which the
computations will be carried out in Sec.~\ref{SecPropEqMain} and
Sec.~\ref{SecGaugFix}. 
	
\subsection{Rank 1}
\label{SubSecRk1}
	
The conformally invariant equation for the  vector field $A^\mu$ is
\begin{equation}
	\label{eq.VectGene1} 
	\square A^\mu 
		- \frac{4}{d}\nabla^\mu \nabla\cdot A 
		- \frac{2}{d-2} R^\mu{}_\nu A^\nu
		- \frac{1}{4}\frac{d(d-4)}{(d-1)(d-2)} R A^\mu 
		= 0,
\end{equation}
with $\overline{A}{}^\mu = \omega^{-d/2} A^\mu$
and might be rewritten as
\begin{equation}
	\label{eq.VectGene2}
	\square A^\mu 
    - \frac{4}{d}\nabla_\nu\nabla^\mu A^\nu 
    + \frac{2}{d}\frac{d-4}{d-2} R^\mu{}_\nu A^\nu
    - \frac{1}{4}\frac{d(d-4)}{(d-1)(d-2)} R A^\mu
    = 0,
\end{equation}
to make it obvious that for $d=4$ it reduces to the free Maxwell's equations on 
the vector potential $A$.
Such an equation, for arbitrary $d$, might be found in many instances.\cite{Deser:1983mm}

For $d=4$ Eq.~\eqref{eq.VectGene1} admits $A_\mu = \nabla_\mu\varphi$ as a 
gauge solution with the scalar $\varphi$ unconstrained.
The gauge invariant quantities are components of the field strength $F$ 
\begin{equation}
	\label{TensFaraday}
	(F(A))^{\alpha\,\mu} 
		= \nabla^\alpha A^\mu - \nabla^\mu A^\alpha,
\end{equation}
which is gauge invariant since for $A^\mu = \nabla^\mu \varphi$ one has
$	(F(A))^{\alpha\,\mu}
    = \bigl[\nabla^\alpha, \nabla^\mu\bigr] \varphi
    = 0.
$

It has been demonstrated in both flat and (A)dS spacetimes \cite{Binegar1982Mink,Binegar1982AdS,Bayen:1984dt,Faci2009} that maintaining conformal covariance through a conformally invariant gauge fixing condition when quantizing $A$ results in straightforward and concise expressions.
This article adopts this approach.

Note that Eq.~\eqref{eq.VectGene1} provides a conformally invariant gauge fixing equation for the four dimensional case.
Indeed, considering its divergence leads to
\begin{equation*}
	\left(\frac{d-4}{d}\right)\Bigl[
	\square\nabla\cdot A
	 +\frac{d}{d-2}\Bigl( \nabla^\mu R_{\mu\nu} A^\nu	
	 -\frac{1}{4}\frac{d}{d-1}\nabla_\mu R A^\mu\Bigr)\Bigr] 
	 = 0.
\end{equation*}
For $d\neq 4$ the term within the brackets necessarily vanishes on the space of
solutions of Eq.~\eqref{eq.VectGene1}, then is conformally invariant.
For $d = 4$ the divergence of Eq.~\eqref{eq.VectGene1} vanishes  automatically and
$A$ is unconstrained.
However, the Eastwood-Singer gauge fixing equation,\cite{EastwoodSinger} which is the $d=4$ content within the bracket,
\begin{equation}
	\label{GaugeEastwoodSinger}
	G(A) = \square \nabla\cdot A
		+ 2 \nabla^\mu R_{\mu\nu} A^\nu
		- \frac{2}{3} \nabla_\mu R A^\mu 
		= 0,
\end{equation}
is conformally invariant on the space of solutions of
Eq.~\eqref{eq.VectGene1} since 
\begin{equation}
	\label{VariationES}
	\overline{G}(\overline{A})
	= \omega^{-4} G(A) 
    + 2\omega^{-5} (\square A^\mu - \nabla^\mu\nabla\cdot A - R^\mu{}_\nu A^\nu)
        (\nabla_\mu \omega),
\end{equation}
where Eq.~\eqref{eq.VectGene1} appears in the right hand side.
This gauge fixing equation restricts the gauge freedom as now $\varphi$ has to fulfill the $d=4$ Paneitz equation\cite{Paneitz1983,Paneitz2008}
\begin{equation}
	\label{eqPaneitz}
	\Bigl(\square^2 + 2\nabla^\mu R_{\mu\nu} \nabla^\nu 
    - \frac{2}{3}\nabla_\mu R\nabla^\mu\Bigr)\varphi = 0,
\end{equation}
which is a fourth order conformally invariant equation on scalars of null conformal weight.
This operator has also been found by other means in other contexts.\cite{Riegert1984,Fradkin1984,Antoniadis:1992}

\subsection{Rank 2}
\label{SubSecRk2}

One can try to generalize the previous scheme to the symmetric traceless rank
$2$ tensor $A^{\mu\nu}$.
Then one can check that
\begin{multline}
	\label{eq.T2}
	    \square A^{\mu\nu}
		- \frac{4}{d+2} (\nabla^\mu \nabla_\rho A^{\nu\rho}
            + \nabla^\nu \nabla_\rho A^{\mu\rho}) 	
		+ \frac{8}{d(d+2)} g^{\mu\nu}\nabla_\rho\nabla_\sigma A^{\rho\sigma}
		- \frac{(d^2 - 2d + 8)}{4 d (d-1)} R A^{\mu\nu}	
        \\
		+ \frac{2}{d}(R^\mu{}_\rho A^{\nu\rho} + R^\nu{}_\rho A^{\mu\rho})
		- \frac{4(d-1)}{d} R^\mu{}_{\rho\sigma}{}^{\nu} A^{\rho\sigma}
		- \frac{4}{d} g^{\mu\nu} R_{\rho\sigma} A^{\rho\sigma}
        = 0
\end{multline}
is conformally invariant\cite{Bayen1970,Drew:1980yk,Barut:1982nj,Deser:1983mm,Leonovich:1984cf,Achour:2013afa} with $\overline{A}^{\mu\nu} = \omega^{-1-d/2} A^{\mu\nu}$.
However, this equation is not unique due to the conformal invariance of the Weyl tensor
\begin{equation*}
	C^\mu{}_{\rho\nu\sigma}
	= R^\mu{}_{\rho\nu\sigma}
    - \frac{1}{d-2}\left(\delta^\mu_{[\nu}R_{\sigma]\rho} 
        - g_{\rho[\nu}R_{\sigma]}^\mu\right) 
    + \frac{1}{(d-1)(d-2)} R \delta^\mu_{[\nu}g_{\sigma]\rho}
\end{equation*}
since one can add a term such as $+ \lambda C^\mu{}_{\rho\sigma}{}^\nu A^{\rho\sigma}$,
where $\lambda$ is unconstrained by the requirement that the resulting equation
has to be Weyl invariant. 

To reduce this freedom with respect to the Weyl invariance of Eq.~\eqref{eq.T2}
we will narrow the scope of this article to conformally
flat Einstein spacetimes (CFES), for which the Riemann and Ricci tensors are 
expressed as
\begin{subequations}
    \label{eq:CFESconstraints}
    \begin{align}\label{RiemT}
        &R_{\mu\nu\rho\sigma} 
            = \frac{R}{d(d-1)}(g_{\mu\rho}g_{\nu\sigma}
                - g_{\mu\sigma}g_{\nu\rho}), \\
        &R_{\mu\nu} 
            = \frac{R}{d} g_{\mu\nu},  \label{RicciConst}\\
        &R = \text{Const}. \label{ScalCurv}
    \end{align}
\end{subequations}
Examples of CFES are Minkowski and (A)dS spacetimes.
Then, using Eq.~\eqref{eq:CFESconstraints}, Eq.~\eqref{eq.T2}
reduces to
\begin{multline}
	\label{Rk2EqCC}
        (E_2(A))^{\mu\nu}
        = \square A^{\mu\nu}
        - \frac{(d^2 - 2d + 8)}{4 d (d-1)} R A^{\mu\nu}
        - \frac{4}{d+2} (\nabla^\mu \nabla_\rho A^{\nu\rho} 
        + \nabla^\nu \nabla_\rho A^{\mu\rho})\\
        + \frac{8}{d(d+2)} g^{\mu\nu}\nabla_\rho\nabla_\sigma A^{\rho\sigma}
        = 0.
\end{multline}
Now, considering the field $A$ as the derivation of a vector field $V$ such as
\begin{equation}
	\label{Rk2fromRk1}
	A^{\mu\nu} 
        = \nabla^\mu V^\nu 
        + \nabla^\nu V^\mu 
        - \frac{2}{d} g^{\mu\nu} \nabla\cdot V,
\end{equation}
then, after a computation involving Eqs.~\eqref{eq:CFESconstraints}, one gets
\begin{equation}
	\label{EqRk2toRk1}
	(E_2(A))^{\mu\nu} = \Bigl(\frac{d-2}{d+2}\Bigr) \Bigl[ 
        \nabla^\mu (E_1(V))^\nu + \nabla^\nu (E_1(V))^\mu 
        - \frac{2}{d} g^{\mu\nu} \nabla_\rho (E_1(V))^\rho\Bigr].
\end{equation}
Similarly, suppose in Eq.~\eqref{Rk2fromRk1} that
$V^\mu = \nabla^\mu\varphi$, then simplifying Eq.~\eqref{EqRk2toRk1} leads to
\begin{equation*}
    (E_2(A))^{\mu\nu} 
        = \frac{(d-2)(d-4)}{d(d+2)}
        \Bigl(
            \nabla^\mu\nabla^\nu + \nabla^\nu\nabla^\mu 
            - \frac{2}{d}g^{\mu\nu}\square\Bigr) E_0(\varphi),
\end{equation*}
meaning that Eq.~\eqref{Rk2EqCC} for $d=4$ allows the gauge freedom up to a scalar 
\begin{equation*}
      A_{\mu\nu} \mapsto {}^\varphi A_{\mu\nu} 
      = A_{\mu\nu} 
      + \Bigl(\nabla_\mu\nabla_\nu 
      - \frac{1}{d}g_{\mu\nu} \square\Bigr) \varphi.
\end{equation*}

Regarding this gauge freedom one can search for a gauge condition similar to the Eastwood-Singer equation.
Taking the divergence of Eq.~\eqref{Rk2EqCC} once and twice yields
\begin{align*}
	&\nabla_\mu (E_2(A))^{\mu\nu} 
    = \Bigl(\frac{d-2}{d+2}\Bigr) (E_1(\nabla\cdot A))^\nu,\\
    &\nabla_\nu\nabla_\mu (E_2(A))^{\mu\nu}
    = \frac{(d-2)(d-4)}{d(d+2)} E_0(\phi),
\end{align*}
where $(\nabla\cdot A)^\nu = \nabla_\mu A^{\mu\nu}$ and  $\phi = \nabla_\mu\nabla_\nu A^{\mu\nu}$.
Those two results then suggest that the set 
\begin{equation*}
	\left\{\begin{aligned}
		& E_2(A) = 0, \\
		& E_0(\phi ) =0,
		\end{aligned}\right.
	\quad (d=4)
\end{equation*}
is conformally invariant, while restricting the gauge freedom allowed by
$E_2(A)=0$.
The conformal invariance is indeed preserved, see Sec.~\ref{SecGaugFix},
and the scalar field $\varphi$ now has to fulfill
\begin{equation*}
	\Bigl(\square - \frac{1}{6}R\Bigr) \square 
    \Bigl(\square + \frac{1}{3}R\Bigr) \varphi = 0.
\end{equation*}
Such a conformally gauge fixing equation was found already
on flat spacetime\cite{DelfinoGalles:1985bb} for the tracefull field,
it reads:
$(\square\partial_\mu\partial_\nu - \frac{1}{4}\eta_{\mu\nu}\square^2)A^{\mu\nu}
= 0$.

\section{Properties of Eq.~(1)}
\label{SecPropEqMain}

\subsection{Conformal invariance on arbitrary spacetimes}

\begin{property}
    The equation\cite{Wunsch86} 
    \begin{align}
        (\square + c_s R) A^{\mu_1\dots\mu_s}
    		&+ a_s s\  \nabla^{(\mu_1} \nabla_{\sigma} A^{\mu_2\dots\mu_s)\sigma}
                \nonumber\\
    		&+ b_s s(s-1)\ g^{(\mu_1\mu_2} \nabla_{\rho}\nabla_{\sigma}
                A^{\mu_3\dots\mu_s)\rho\sigma} \nonumber\\
    		&+ d_s s R^{(\mu_1}{}_{\sigma} A^{\mu_2\dots\mu_s)\sigma}
                \label{EqWeylTotale}\\
    		&+ e_s s(s-1)\  R^{(\mu_1}{}_{\rho\sigma}{}^{\mu_2}
                A^{\mu_3\dots\mu_s)\rho\sigma} \nonumber\\
    		&+ f_s s(s-1)\  R_{\rho\sigma} g^{(\mu_1\mu_2}
                A^{\mu_3\dots\mu_s)\rho\sigma} \nonumber
        = 0,
    \end{align}
    where the conformal weight is $h(A^{\mu_1\dots\mu_s}) = 1 - s - d/2$,
    the coefficients $a_s$, $b_s$ and $c_s$ are given in Eq.~\eqref{Coefficients} and 
    \begin{equation*}
    	d_s = \frac{2}{d},\quad
    	e_s = -\frac{2}{d}\Bigl(\frac{d-1}{s-1}\Bigr),\quad
    	f_s = -\frac{2}{d(s-1)}\Bigl(\frac{d+s-2}{d +2s-4}\Bigr),
    \end{equation*}
    is conformally invariant under generic Weyl rescalings.
\end{property}
\begin{proof}
    This is verified by a direct brute-force computation using Eqs.~\eqref{ORiemann}
    and \eqref{ORicci}.
\end{proof}

\begin{remark}
    As noticed in Sec.~\ref{SubSecRk2} Eq.~\eqref{EqWeylTotale} is not unique as one can add a term such as
\begin{equation}
	\label{chg.eqtotale}
	+ \lambda s(s-1)\ C^{(\mu_1}{}_{\rho\sigma}{}^{\mu_2}
        A^{\mu_3\dots\mu_s)\rho\sigma} ,
\end{equation}
that is, changing in Eq.~\eqref{EqWeylTotale} the coefficients according to 
\begin{align*}
    &c_s \mapsto c_s + \frac{\lambda s (s-1)}{(d-1)(d-2)},
    &&d_s \mapsto d_s - 2\lambda \left(\frac{s-1}{d-2}\right),\\
    &e_s \mapsto e_s + \lambda,
    &&f_s \mapsto f_s + \frac{\lambda}{d-2},
\end{align*}
while keeping the Weyl invariance of the resulting equation intact.
\end{remark}

\begin{corollary}
    Equation~\eqref{EqPrinc} being the restriction of Eq.~\eqref{EqWeylTotale} to CFES it inherits its (restricted) conformal invariance
\end{corollary}

\subsection{Gauge invariance at $d=4$}
\label{SubSecGI}

\begin{definition}
    For a rank $s\geq 2$, let us define the symmetric traceless gradient 
    ($\STG$):
    \begin{equation}
        \label{defSTG}
    	(\STG(f))^{\mu_1\dots\mu_s} 
        = s \nabla^{(\mu_1} f^{\mu_2\dots\mu_s)}
            - \frac{s(s-1)}{(d + 2s-4)} g^{(\mu_1\mu_2} \nabla_{\sigma}
                f^{\mu_3\dots\mu_s)\sigma},
    \end{equation}
    with $f$ a symmetric traceless field of rank $s-1$ and $\STG(f)$ a symmetric traceless field of rank $s$.
    In addition, let us commit the abuse of language $(\STG(\varphi))^\mu = \nabla^\mu \varphi$, for $\varphi$ a scalar field.
\end{definition}

\begin{property}
    Equation~\eqref{EqPrinc} is gauge invariant (only) for $d=4$ with the gauge freedom up to a scalar
    \begin{equation}
    	\label{GaugeInv}
    	A \mapsto {}^\varphi A = A + \STG^s(\varphi).
    \end{equation}
\end{property}

\begin{proof}
    First note that, using the symmetric traceless gradient, Eq.~\eqref{EqPrinc} might be rewritten as
    \begin{equation}
        \label{EqPrincSTG}
    	(\square + c_s R) A + a_s \STG(\nabla\cdot A) = 0.
    \end{equation}
    Second, let us consider the field $A = \STG(f)$.
    Since the coefficients $(a_s, c_s)$ fulfill the recurrence relations
    \begin{subequations}
        \label{eq:RecSTG}
        \begin{align}
            a_{s-1} &= \frac{a_s}{1+a_s}\Bigl(\frac{d + 2s-6}{d+2s-4}\Bigr), \\
            c_{s-1} &= \frac{1}{1 + a_s}\Bigl( \frac{d + 2s - 3}{d(d-1)}
                + \frac{(s-1)(d+s-3)}{d(d-1)} a_s + c_s\Bigr),
        \end{align}
    \end{subequations}
one gets the identity
\begin{equation}
	\label{ID.STG1}
	E_s(\STG(f)) = (1+a_s) \STG(E_{s-1}(f)).
\end{equation}

Now Eq.~\eqref{ID.STG1} enables us to look for gauge solutions obtained from a field $g$ of rank $r$, with $r < s$, as one obtains
\begin{equation}
	\label{ID.STG2}
	E_s(\STG^{s-r}(g))
    = \prod_{i=r+1}^s (1+a_i)\ \STG^{s-r}(E_r(g)).
\end{equation}
Then, the existence of gauge solutions amounts to look if there is a rank $r$ and a dimension $d$ such that 
$1+a_r = 0$.
From the values of the $a_i$'s, given in Eq.~\eqref{Coefficients},
there is only one (physical) solution given by $(a_1, d = 4)$ corresponding to the gauge freedom up to a scalar.
\end{proof}

\section{The conformally invariant gauge fixing equation}
    \label{SecGaugFix}

\subsection{Uncovering the gauge fixing equation}
	\label{SubSecDerivGFE}

Taking the divergence of Eq.~\eqref{EqPrinc} yields
\begin{equation}
	\label{ID.Div1}
	\nabla_{\mu_s} (E_s(A))^{\mu_1\dots\mu_s}
    = (1+a_s)(E_{s-1}(\nabla\cdot A))^{\mu_1\dots\mu_{s-1}},
\end{equation}
through a direct computation, involving the commutation of covariant derivatives, using the fact that the  Riemann tensor is given by Eq.~\eqref{RiemT} with
$R = \text{Const}.$ and  that the coefficients $(a_s, b_s, c_s)$ are solutions
of the recurrence relations
\begin{subequations}
    \label{eq:Rec}
    \begin{align}
         a_{s-1} & = \frac{a_s + 2 b_s}{1+ a_s}, \\
         b_{s-1} & = \frac{b_s}{1+a_s}, \\
         c_{s-1} & = \frac{1}{1+a_s}\Bigl(\frac{d + 2s - 3}{d(d-1)}
            + \frac{(s-1)(d+s-3)}{d(d-1)} a_s + c_s\Bigr).
    \end{align}
\end{subequations}
That is, the $1$-divergence of $A$ satisfies the equation of a rank $s-1$ symmetric
traceless field.
By induction each divergence has to fulfill the equation $E_i$ of its rank and finally
\begin{equation}
    \label{ID.Div2}
	\nabla_{\mu_1}\dots\nabla_{\mu_s} (E_s(A))^{\mu_1\dots\mu_s}
    = \prod_{i=1}^s (1+a_i)\ E_0(\phi ).
\end{equation}
Therefore, the behaviour of the divergences of $A$ is determined by
Eq.~\eqref{EqPrinc}.
That is, if none of the pre-factor $(1+a_i)$ vanish.
Identically to Sec.~\ref{SubSecGI} for $d=4$ the prefactor $(1+a_1)$ vanishes.
This is the confirmation of the gauge freedom up to a scalar shown before for which
\begin{equation*}
    \phi = \nabla_{\mu_1}\dots\nabla_{\mu_s} A^{\mu_1\dots\mu_s}, \quad (d=4)
\end{equation*}
is left free by Eq.~\eqref{EqPrinc}.
This should not come as a surprise as if one sets $b_s = -a_s/(d+2s-4)$, to make obvious the symmetric traceless gradient in Eq.~\eqref{EqPrinc}, then the recurrence relations in Eq.~\eqref{eq:Rec} become those of Eq.~\eqref{eq:RecSTG}. 

To conclude, we have shown that the $s$-fold divergence $\phi$ is unconstrained by
Eq.~\eqref{EqPrinc} if $d=4$.
In this context, enforcing $E_0(\phi) = (\square -\frac{1}{6}R)\phi = 0$ as a gauge fixing equation seems justified in terms of conformal invariance.
This is because, firstly, solutions to Eq.~\eqref{EqPrinc} that also satisfy Eq.~\eqref{EQFG} are left invariant under conformal transformations (as we will demonstrate), and secondly, in arbitrary dimensions where $d \neq 4$, the associated equation is inherently satisfied by the solutions of Eq.~\eqref{EqPrinc}.
When $d=4$, this discrepancy is resolved by applying the gauge fixing equation.
 
\subsection{Restricted Weyl invariance of the gauge fixing equation}

As this work is concerned with CFES we are interested in
rescalings mapping a CFES on another.\cite{Brinkmann}
A smaller class of $\omega$'s has to be considered, this motivates the following definition.

\begin{definition}
    A Weyl transformation will be called a restricted Weyl transformation if
    $\omega$ fulfills the two following equations
    \begin{align}
    	\label{WeylRestr}
    	&\Bigl(\nabla_\mu\nabla_\nu 
            - \frac{1}{d}g_{\mu\nu}\square \Bigr)\frac{1}{\omega} 
        = 0,\\
        \label{WeylRestr1}
    	&\square\nabla_\mu \omega 
        = (\nabla_\mu\omega)
            \Bigl( 3 \omega^{-1} (\square\omega) - \frac{R}{d(d-1)}\Bigr)
    	    - (d-4)\, \omega^3 \Bigl(\nabla^\alpha\frac{1}{\omega}\Bigr)
            \Bigl(\nabla_\mu\nabla_\alpha\frac{1}{\omega}\Bigr).
    \end{align}
\end{definition}
\begin{property}
    Restricted Weyl transformations map CFES on CFES.
\end{property}
\begin{proof}
Asking for a rescaling to map a CFES to another CFES is asking for 
the relations \eqref{eq:CFESconstraints} to be preserved.
First, consider Eq.~\eqref{RicciConst} on $(M,\overline{g})$, that is
\begin{equation}
	\label{mapConst}
	\overline{R}_{\mu\nu} = \frac{1}{d}\overline{R}\ \overline{g}_{\mu\nu}.
\end{equation}
Then, plugging Eq.~\eqref{ORicci} in the left-hand side of Eqs.~\eqref{mapConst}
and \eqref{OCurv} in the right-hand side and finally, since $(M,g)$ is
also a  CFES, using Eq.~\eqref{RicciConst} to further reduce the equality one
obtains that Eq.~\eqref{WeylRestr} has to be fulfilled.

In addition, the Weyl rescaled curvature scalar $\overline{R}$ has to be constant.
Considering the equation $\overline{\nabla}_\mu \overline{R} = 0$ and using Eq.~\eqref{ScalCurv} and, since $R$ is also constant, one gets that Eq.~\eqref{WeylRestr1} has to be fulfilled, with the right hand side tweaked for later convenience.
\end{proof}

\begin{property}\label{prop:Restricted_GF}
    Over four dimensional CFES the gauge fixing equation
    \begin{equation}
    	\label{EQFG}
        E_0(\phi) = \Bigl(\square - \frac{1}{6} R\Bigr)\phi = 0,
    	\quad (d=4)
    \end{equation}
    with $\phi = \nabla_{\mu_1}\dots\nabla_{\mu_s} A^{\mu_1\dots\mu_s}$,
    is invariant under restricted Weyl transformations on the space of solutions of Eq.~\eqref{EqPrinc}.
\end{property}

Due to its length the proof of Prop.~\ref{prop:Restricted_GF} has been moved in App.~\ref{app:Restricted_GF}.

\section{The field strength $F$}
\label{SecFS}

Equation~\eqref{EqPrinc} admits gauge solutions through the gauge transformation Eq.~\eqref{GaugeInv}.
In this section we construct a field strength $F$ upon the potential $A$  by proceeding in close analogy with Maxwell's case.
First, writing the field equation as a divergence provides a candidate $F$ for the field strength, its symmetries are registered.
Then, we show on CFES that $F$ vanishes on gauge solutions, hence \emph{is} a field strength.
The equations on $A$ are then translated into a system of two first-order
conformally invariant equations on $F$, one of which vanishes for $F$ derived
from a potential and $(M,g)$ being conformally flat.

\subsection{Constructing the field strength $F$} 

To produce a field strength let us consider instead of Eq.~\eqref{EqPrinc} the
equation
\begin{equation}
    \label{eq.alternat1}
    \square A - \frac{(d+2s-4)}{s(d+s-3)}\STG (\nabla\cdot A)
    - \frac{(d+s-2)}{d(d-1)} R A = 0,
\end{equation}
which in arbitrary dimension $d$ has the gauge invariance up to a scalar field showed in
Eq.~\eqref{GaugeInv} since its coefficients also fulfill the recurrence relations~\eqref{eq:RecSTG} and $(1+a_1) = 0$.
For $s=1$ Eq.~\eqref{eq.alternat1} is the restriction of Maxwell equation
$\nabla_\mu(\nabla^\mu A^\nu - \nabla^\nu A^\mu) = 0$ to a CFES.
For $d=4$ Eq.~\eqref{EqPrinc} and Eq.~\eqref{eq.alternat1} are the same.

\begin{definition}
    Let $A$ be a symmetric traceless field of rank $s$, let us define the field strength $F$ built upon $A$ by
    \begin{equation}
    	\label{FStrF}
        \begin{split}
            (F(A))^{\alpha\ \mu_1\dots\mu_s} 
            & = (\mathcal{D} A)^{\alpha\ \mu_1\dots\mu_s} \\
            & =  \nabla^\alpha A^{\mu_1\dots\mu_s} 
                -  \nabla^{(\mu_1} A^{\mu_2\dots\mu_s) \alpha} \\
            &\quad
                - \frac{(s-1)}{(d+s-3)} \bigl[
                    g^{\alpha (\mu_1} \nabla_{\sigma} A^{\mu_2\dots\mu_s) \sigma}
                    - g^{(\mu_1\mu_2} \nabla_\sigma
                        A^{\mu_3\dots\mu_s) \alpha\sigma}
                \bigr], 
        \end{split}
    \end{equation}
    found already in flat space in Ref.~\onlinecite{Erdmenger2} and from which we borrow the notation
    $\mathcal{D}$.
    For $s=1$ one has $F^{\alpha\, \mu} = \nabla^\alpha A^\mu - \nabla^\mu A^\alpha$.
\end{definition}

Under the assumption that the underlying space is a CFES,  for which Eq.~\eqref{eq:CFESconstraints} is fulfilled,
Eq.~\eqref{eq.alternat1} can be recast in terms of $F$ as
\begin{equation}
	\label{eq.alternat2}
	\nabla_\alpha (F (A))^{\alpha\ \mu_1\dots\mu_s}= 0.
\end{equation}
\begin{remark}
    On a generic spacetime the equation resulting from
    \eqref{eq.alternat2} is still conformally invariant for $d=4$ as one schematically gets
    \begin{equation*}
    	\nabla_\alpha (F(A))^{\alpha\ \mu_1\dots\mu_s} 
    	= \eqref{EqWeylTotale} + \eqref{chg.eqtotale}, \quad
    	\lambda = \frac{s+2}{s(s-1)}, \quad
    	d=4.
    \end{equation*}
\end{remark}
	
From the definition of $F$, see Eq.~\eqref{FStrF}, one can workout the identities fulfilled by the field strength, namely
\begin{equation}\label{ID.FS}
    F^{(\alpha\ \mu_1\dots\mu_s)}= 0, \qquad
    g_{\mu_i\mu_j}  F^{\alpha\ \mu_1\dots\mu_s} =0, \qquad
    g_{\alpha\mu_i} F^{\alpha\ \mu_1\dots\mu_s} =0,
\end{equation}
thanks to which one can find that $F$ possesses
\begin{equation}
	\label{dofFp}
	\text{DoF}(F) = \frac{(d+s-4)!}{(d-3)!(s+1)!} s (d + 2s -2)(d+s-2)
\end{equation}
independent components.
Similarly to the Faraday tensor the relations \eqref{ID.FS} can serve
as the defining properties of $F$ and under the appropriate assumptions
eventually there would exist a potential $A$ such that $F$ is given by
Eq.~\eqref{FStrF}.

Finally, we show that $F$ deserves its label of field strength, that is: it has to be gauge invariant.

\begin{property}
    On a CFES $F$ is gauge invariant, namely $F(\STG^s(\varphi)) = 0$ with $\varphi$ an arbitrary scalar field.
\end{property}

\begin{proof}
Let $(M,g)$ be a CFES and consider the field $A = \STG(f)$.
Let us also make the extra assumption that the field $f$ satisfies
Eq.~\eqref{eq.alternat1} written at the rank $s-1$.
Then, after some algebra in which the strategy is that as soon as the
laplacian acts on $f$ to replace it by the remaining of Eq.~\eqref{eq.alternat1},
one gets
\begin{align*}
    (F(\STG (f)))^{\alpha\ \mu_1\dots\mu_s}
    &= (s-1) \Bigl[ 
        \nabla^{(\mu_1} (F(f))^{|\alpha |\ \mu_2\dots\mu_s)}
    - \frac{s-1}{d + 2s-4} g^{(\mu_1\mu_2} \nabla_{\sigma}
                    (F(f))^{|\alpha |\ \mu_3\dots\mu_s) } \Bigr] \\
    &= \Bigl(\frac{s-1}{s}\Bigr) 
        (\STG (F(f)))^{\lvert\alpha\rvert\ \mu_1\dots\mu_s},
\end{align*}
in which the $\lvert\alpha\rvert$ means that this superscript $\alpha$ is not
involved in the $\STG$ process.
Suppressing the indices it reads as
\begin{equation}
	\label{ID.FS4}
	F(\STG (f)) = \Bigl(\frac{s-1}{s}\Bigr) \STG (F(f)).
\end{equation}

This scheme can be pushed further.
Let $g$ be a symmetric traceless field of rank $r$, with $r < s$, solution of
Eq.~\eqref{eq.alternat1} written at the rank $r$.
Using Eq.~\eqref{ID.STG2} adapted to Eq.~\eqref{eq.alternat1} it follows that
$\STG^{n}(g)$ is a solution of Eq.~\eqref{eq.alternat1} written at the
rank $r+n$.
Then, from Eq.~\eqref{ID.FS4} one gets
\begin{equation}
	\label{ID.FS5}
	F(\STG^{s-r}(g)) = \frac{r}{s} \STG^{s-r}(F(g)).
\end{equation}
Finally, one can consider a scalar field $\varphi$ and a pure gauge field 
obtained from that scalar $A = \STG^s(\varphi)$.
Since Eq.~\eqref{eq.alternat1} has solutions determined up to a scalar
$\STG^n(\varphi)$ is a solution whatever the rank is and Eq.~\eqref{ID.FS5} reads as
\begin{equation*}
	F(\STG^s(\varphi))
    = \frac{1}{s} \STG^{s-1}(F(\STG(\varphi ))
    = 0,
\end{equation*}
in which the last equality is nothing but $\nabla^{[\alpha}\nabla^{\mu]}\varphi = 0$.
Thus, the field strength $F$ vanishes on gauge solutions.
\end{proof}

\begin{remark}
The gauge invariance of $\mathcal{D}A$ has been previously recognized in flat space.\cite{Erdmenger2}
For extending this concept to more general curved backgrounds (than CFES), a deeper insight into the gauge solutions would be necessary.
\end{remark}

\subsection{Conformal invariance of the equations fulfilled by $F$ at $d=4$}
\label{EquationsCompletes}

Here we show that one can find a set of first order equations fulfilled by $F$ which are conformally invariant for $d=4$.

\begin{definition}
Let us assume that $F$ fulfills Eq.~\eqref{ID.FS}, with no further assumptions,
and let us define
\begin{align}
    (\mathcal{D}F)^{\alpha\beta\ \mu_1\dots\mu_s}
     &= \nabla^{[\alpha}F^{\beta]\ \mu_1\dots\mu_s}
     + \frac{s}{s+1} \nabla^{(\mu_1}F^{|[\alpha\ \beta]|\mu_2\dots\mu_s)}
    \notag\\
    &\quad\label{Def.DFp}
    - \frac{s}{(d+s-4)(s+1)} \Bigl[
          s\, g^{(\mu_1 \lvert [\alpha}\nabla_{\sigma}
            F^{\beta]\rvert\ \mu_2\dots\mu_s) \sigma} \\ &\qquad\qquad\notag
        + (s-1)  g^{(\mu_1\mu_2} \nabla_{\sigma}
            F^{\lvert[\alpha\ \beta]\rvert\mu_3\dots\mu_s )\sigma}
        + g^{(\mu_1 \lvert [\alpha}\nabla_{\sigma}
            F^{\lvert\sigma\rvert\  \beta]\rvert\mu_2\dots\mu_s)}
    \Bigr], 
\end{align}
in which keeping the notation $\mathcal{D}$ seems to be natural if one compares
Eq.~\eqref{Def.DFp} and Eq.~\eqref{FStrF}, thus generalizing $\mathcal{D}$ to higher ranks and on curved spaces.
\end{definition}

\begin{remark}
Note that for  $s=1$ one gets
\begin{align*}
    (\mathcal{D}F)^{\alpha\beta\ \mu}
    &= \nabla^{\alpha}F^{\beta\ \mu} 
    - \nabla^{\beta}F^{\alpha\ \mu}
    + \frac{1}{2} \nabla^\mu F^{[\alpha\ \beta]} \\
    &= \nabla^{\alpha}F^{\beta\ \mu} 
    + \nabla^{\beta}F^{\mu\ \alpha}
    +  \nabla^\mu F^{\alpha\ \beta}.
\end{align*}
\end{remark}

Thanks to the differential operator $\mathcal{D}$ we can consider for $F$ a system of first order differential equations.

\begin{property}
    The system 
    \begin{equation}
        \label{EqMaxwellcomplete}
        \left\{
        \begin{aligned}
            &(\mathcal{D}F)^{\alpha\beta\ \mu_1\dots\mu_s}  = 0, \\
            &\nabla_{\alpha}F^{\alpha\ \mu_1\dots\mu_s} = 0,
        \end{aligned}\right.
    \end{equation}
    is conformally invariant with $h(F^{\alpha\ \mu_1\dots\mu_s}) = -3- s$ and if, and only if, $d=4$.
\end{property}
\begin{proof}
    Direct computation on $\overline{\mathcal{D}}\,\overline{F}$ since the second half is already known to be conformally invariant if, and only if, $d=4$.
\end{proof}

\begin{remark}
This has to be compatible with the case in which $F=\mathcal{D}A$ as in Eq.~\eqref{FStrF}. One gets
\begin{align*}
    (\mathcal{D}F)^{\alpha\beta\ \mu_1\dots\mu_s}
    &=(\mathcal{D}^2 A)^{\alpha\beta\ \mu_1\dots\mu_s} \\
    &=(1-s)\left[
        C^{\alpha\beta}{}_{\sigma}{}^{(\mu_1} A^{\mu_2\dots\mu_s)\sigma}  
        +  C_{\sigma}{}^{(\mu_1\mu_2\lvert [\alpha}
            A^{\beta]\rvert\mu_3\dots\mu_s)\sigma} \right] \\
    &\quad
        + \frac{(s-1)(s-2)}{d+s-4}
        C^\rho{}_{\nu_1\nu_2}{}^{\sigma}\times\\
    &\qquad\qquad
    \times\Bigl[
            \delta^{[\alpha}_{\rho} g^{\beta](\mu_1} \delta^{\mu_2}_{\sigma}
            \delta^{\mu_3}_{\nu_3}  
        + \delta^{[\alpha}_{\nu_3} g^{\beta](\mu_1} \delta^{\mu_2}_{\rho}
            \delta^{\mu_3}_{\sigma}
        + \delta^{[\alpha}_{\nu_3} \delta^{\beta]}_{\sigma}
            \delta^{(\mu_1}_{\rho} g^{\mu_2\mu_3} \Bigr]
    A^{\mu_4\dots\mu_s)\nu_1\nu_2\nu_3}
\end{align*}
which vanishes when either $s=1$ or if $(M,g)$ is conformally flat for which one gets
\begin{equation*}
    \mathcal{D}^2 A = 0, \quad (M,g) \equiv \text{conformally flat}.
\end{equation*}
The $d=4$ conformal invariance of $F$ is compatible with that of $A$ with 
$\overline{F}(\overline{A}) = \omega^{-3-s}F(A)$.
\end{remark}

\section{An electromagnetic decomposition of $F$}
\label{sec:E-B_decomposition}

Note that from Eq.~\eqref{dofFp} that for $d=4$ the following occurs 
\begin{equation*}
    \text{DoF}(F) \Big\rvert_{d=4} 
    = 2 s (s+2) = 2 \sum_{j=1}^s (2 j +1) .
\end{equation*}
In other words, the number of independent components of $F$ corresponds to that of the sum over 
$2s$ symmetric traceless tensors with respect to SO(3).
In this section, we demonstrate that $F$ can indeed be decomposed in a set of such fields, akin to electric and magnetic fields $E$-$B$.
We first define these fields and show that $F$ can be fully expressed in terms of them.
Next, we write how the set of conformally invariant equations on $F$, as given in Eq.~\eqref{EqMaxwellcomplete}, turns out on those
fields, a task simplified once it is shown that \eqref{EqMaxwellcomplete}
have a duality property.
Finally, we register how the algebra $\fo(2,4)$ acts upon those fields.

From now on every computation is performed on flat space.
The results which are obtained here can be lifted to conformally flat spacetimes
thanks to the conformal invariance of the system.

\begin{definition}
    On flat space let us consecutively define
    \begin{subequations}
        \label{def.MN}
        \begin{align}
            M_j^{i_1\dots i_j} &= F^{0\ 0\dots 0 i_1\dots i_j}, \\
            N_j^{i_1\dots i_j} &= j  \varepsilon^{(i_1 \lvert bc}
                F^{b\ c 0\dots 0\rvert i_2\dots i_j)},
        \end{align}
    \end{subequations}
    with $1\leqslant j \leqslant s$ and which are symmetric $3$-fields, and
    \begin{subequations}
        \label{def.EB}
        \begin{align}
            E_j^{i_1\dots i_j} &= M_j^{i_1\dots i_j} - \text{traces}, \\
            B_j^{i_1\dots i_j} &= \frac{1}{s+1}N_j^{i_1\dots i_j} - \text{traces},
        \end{align}
    \end{subequations}
    which are  symmetric traceless $3$-fields, the traces are with respect to the $3$
    dimensional metric $\delta_{ij} = \delta^{ij} = -\eta_{ij}$.
\end{definition}

\begin{property}\label{prop:InversionEB}
    The field strength $F$ uniquely decomposes over the $E_j$-$B_j$ fields, that is
    \begin{equation}
        F = \bigoplus_{j=1}^s (E_j\oplus B_j).
    \end{equation}
\end{property}

The field strength $F$ decomposes over the $E_j$-$B_j$'s if an arbitrary
component can, unequivocally, be written in terms of such $3$-dimensional
fields, that is if one can invert Eqs.~\eqref{def.MN} and \eqref{def.EB}.
This, quite lengthy, proof is provided in App.~\ref{InversionFEB}.

Given that $F$ satisfies a set of conformally invariant equations, see Eq.~\eqref{EqMaxwellcomplete}, and considering that $F$ can be broken down into $E$-$B$ fields, it is important to express these equations in terms of those fields.
To carry this out, it is beneficial to initially recognize that these equations possess a duality property.

\begin{property}\label{prop:Ftilde}
    Let\ $\widetilde{\cdot}$\ be the map defined on $F$ as
    \begin{equation*}
        \widetilde{F}^{\alpha\ \mu_1\dots\mu_s}
        = \frac{s}{s+1} \varepsilon^{\alpha(\mu_1}{}_{\beta\gamma} 
            F^{\lvert\beta\rvert\ \mu_2\dots\mu_s)\gamma},
    \end{equation*}
    then $\widetilde{F}$ fulfills the same symmetries as $F$ and one has
    \begin{equation*}
        \widetilde{E}_j = - B_j, \qquad
        \widetilde{B}_j =  E_j,
    \end{equation*}
    and consequently
    \begin{equation*}
        \smash{\widetilde{\widetilde{F}}} = -F.
    \end{equation*}
\end{property}
\begin{proof}
This comes from a careful inspection of $\widetilde{F}$ noticing that it fulfills all the symmetries that $F$ possesses [registered in Eq.~\eqref{ID.FS}].
Then, from these symmetries, $\widetilde{F}$ might be decomposed over fields $\widetilde{E}_j$ and $\widetilde{B}_j$ according to Eqs.~\eqref{def.MN} and \eqref{def.EB}.
Performing one such decomposition, while knowing the $E_j$'s and $B_j$'s of $F$,
one recognizes that
\begin{equation*}
    \widetilde{E}_j = - B_j, \qquad
    \widetilde{B}_j =  E_j,
\end{equation*}
and then $\smash{\widetilde{\widetilde{F}}} = -F$.
\end{proof}

Now, noticing that $*\mathcal{D}F \propto \partial\widetilde{F}$, 
the conformally invariant equations~\eqref{EqMaxwellcomplete} fulfilled by $F$ can be written as
\begin{equation}
    \label{eqMaxwell4}
    \left\{
    \begin{aligned}
        \partial_\alpha \widetilde{F}^{\alpha\ \mu_1\dots\mu_s} &=0, \\
        \partial_\alpha F^{\alpha\ \mu_1\dots\mu_s} &=0,
    \end{aligned}\right.
    \quad \text{ on } \quad (\mathbb{R}^4, \eta).
\end{equation}
From property~\ref{prop:Ftilde} (above) Eq.~\eqref{eqMaxwell4} remain invariant under the electromagnetic duality
\begin{equation}
    \label{dualityEB}
    (E_j, B_j) \mapsto (-B_j, E_j).
\end{equation}
Finally, introducing the notations
\begin{align*}
    &\div E_j \equiv \partial_{i_j} E_j^{i_1\dots i_j},\\
    &\grad E_{j-1}
        \equiv 
        j \delta^{k(i_1}\partial_k E_{j-1}^{i_2\dots i_j)} 
        - \text{traces}, \\
    &\curl E_j
        \equiv
        \varepsilon^{kl (i_1} \partial_k E_j^{i_2\dots i_j)l},
\end{align*}
setting $H_j = E_j + \mathrm{i} B_j$, with $\mathrm{i}^2 = -1$, and leveraging the duality property~\eqref{dualityEB}, Eq.~\eqref{eqMaxwell4} can be recast as
\begin{equation}\label{eq:Maxwell5}
      \partial_t H_j - \mathrm{i} \Bigl(\frac{s+1}{j+1}\Bigr) \curl H_j
	 - \Big(\frac{s-j}{j+1}\Big) \div H_{j+1}
	 + \Bigl(\frac{j^2 + j + 2(s+1)}{j(j+1)(j+2)}\Bigr) \grad H_{j-1}
     = 0, 
\end{equation}
with $H_j = 0$ for $j \leqslant 0$ and for $j > s$.
The complex conjugated equation holds on the complex conjugated fields
$\overline{H}_j$.

On $H_j$ and $\overline{H}_j$ the action of the conformal algebra $\fo(2,4)$ algebra is deduced from that on the field $F$ and it yields
\begin{align*}
    (P_\mu H_j)^{i_1\dots i_j}
        &= \partial_\mu H_j^{i_1\dots i_j}, \\
    (D H_j)^{i_1\dots i_j}
        &= (x\cdot \partial +2) H_j^{i_1\dots i_j}, \\
        (X_{mn} H_j)^{i_1\dots i_j}
        &= x_{[m}\partial_{n]} H_j^{i_1\dots i_j} 
            + (\Sigma_{mn} H_j)^{i_1\dots i_j}
        = x_{[m}\partial_{n]} H_j^{i_1\dots i_j}
            + j \delta_{[m}^k\delta_{n]}^{(i_1} H_j^{i_2\dots i_j)k}, \\
    (X_{0m} H_j)^{i_1\dots i_j}
        &= x_{[0}\partial_{m]}H_j^{i_1\dots i_j}
            + (\Sigma_{0m} H_j)^{i_1\dots i_j} \\
        &= x_{[0}\partial_{m]} H_j^{i_1\dots i_j}
            + \mathrm{i} \Bigl(\frac{s+1}{j+1}\Bigr) \varepsilon^{k m (i_1}
                H_j^{i_2\dots i_j) k}
            - \frac{j(s-j)}{j+1} H_{j+1}^{i_1\dots i_j m}\\&\quad
            - \frac{[(s+3)j(j+1) +2(s+1)]}{(j+1)(j+2)}
            \Bigl(
                \delta^{m (i_1}_{} H_{j-1}^{i_2\dots i_j)}
                - \frac{j-1}{2j-1} \delta^{(i_1i_2}_{}
                    H_{j-1}^{i_3\dots i_j) m}
            \Bigr), \\
    (K_\mu H_j)^{i_1\dots i_j}
        &= x^\nu (X_{\nu\mu} H_j)^{i_1\dots i_j}
            - x_\mu (D H_j)^{i_1\dots i_j}
            - 2 x_\mu H_j^{i_1\dots i_j},
\end{align*}
with $P_\mu$, $D$, $X_{mn}$, $X_{0m}$, $K_\mu$ the generators of the translations, dilations, rotations, Lorentz' boosts and special conformal transformations, respectively.
Note that the full conformal group acts independently on the fields $H_j$ and
$\overline{H}_j$.

\section{Concluding remarks: An action and a scalar product out of $F$}
\label{sec:Conclusion}

Let us conclude with two remarks.
First, note that the equation of motion, Eq.~\eqref{eq.alternat2}, can be
derived from the action
\begin{equation}
    \label{eq:Lagrangien}
    \begin{aligned}
    S[g, A]
        &= \int \frac{s}{2(s+1)} (F(A))_{\alpha\, \mu_1\dots\mu_s} 
            (F(A))^{\alpha\, \mu_1\dots\mu_s} \;
            \mathrm{d}\text{Vol}_g \\
        &=\int  \frac{s}{2(s+1)} (F(A))^2 \; \mathrm{d}\text{Vol}_g.
    \end{aligned}
\end{equation}

From Eq.~\eqref{ID.FS} it is evident that the elements on the ``diagonal'' of $F$ (such as $F^{0\, 0...0}$, $F^{1\, 1...1}$, etc.) are null.
Subsequently, in a coordinate system where the $0$'th coordinate foliates the spacetime by Cauchy hypersurfaces, it follows that the zeroth momenta becomes identically zero in conventional canonical quantization.
This phenomenon suggests the presence of gauge invariance across a class of spacetimes broader than CFES.

Finally, still in view of quantization, thanks to the operator
$\mathcal{D}$ and Eq.~\eqref{eq:Lagrangien}, one can equip the space of
solutions to Eq.~\eqref{EqPrinc} with a symplectic form 
\begin{equation}
    \label{eq:ScalarProduct}
    \langle A, A'\rangle
    = \int \Bigl[ A_{\mu_1\dots\mu_s}(\mathcal{D} A')^{\alpha\
    \mu_1\dots\mu_s}
    - A'_{\mu_1\dots\mu_s}(\mathcal{D} A)^{\alpha\ \mu_1\dots\mu_s}\Bigr]
    \mathrm{d}\Sigma_\alpha,
\end{equation}
in which the integral is carried over the Cauchy hypersurface $\Sigma$.
One such product, for $d=4$, is conformally invariant and vanishes on pure gauge
solutions.

This set the ground for the quantization of fields satisfying  Eq.~\eqref{EqPrinc} when $d=4$ with the help of the conformally invariant gauge fixing equation \eqref{Set1} and the scalar product \eqref{eq:ScalarProduct}.
A task which will be adressed elsewhere.

\appendix

\section{Summary of Weyl rescalings formulas}
\label{SecRemConf}

Starting from a spacetime $(M,g)$ a Weyl rescaling might be defined\cite{WaldRG} as
\begin{equation}
    \label{WeylTransf}
	(M,g)\mapsto (M,\overline{g})
	\quad s.t. \quad
	\overline{g}_{\mu\nu}(x) = \omega^2(x)g_{\mu\nu}(x),
\end{equation}
with $\omega\in C^\infty(M)$.
Then quantities derived from the metric change according to
\begin{align}
	\label{OLeviCiv}
	&\overline{\Gamma}^\rho{}_{\mu\nu} 
	= \Gamma^\rho{}_{\mu\nu} 
		+ \omega^{-1}( 
		\delta_\mu^\rho\delta_\nu^\sigma  
		+\delta_\nu^\rho\delta_\mu^\sigma
		- g_{\mu\nu} g^{\rho\sigma})\omega_{,\sigma}, \\
	&\overline{R}^\mu{}_{\rho\nu\sigma}
	= R^\mu{}_{\rho\nu\sigma}
		-\omega^{-1}(
			\delta_{[\nu}^\mu
			\delta_{\sigma]}^\tau
			\delta_\rho^\varphi
			- g_{\rho[\nu}\delta_{\sigma]}^\tau
			g^{\mu\varphi}) \omega_{;\tau\varphi}
        \notag\\ &\qquad\qquad
		+ \omega^{-2}( 2 \delta_{[\nu}^\mu
			\delta_{\sigma]}^\tau\delta_\rho^\varphi
			- 2 g_{\rho[\nu}\delta_{\sigma]}^\tau
			g^{\mu\varphi}
			+ g_{\rho[\nu} \delta_{\sigma]}^\mu
			g^{\tau\varphi}_{}) \omega_{,\tau}\omega_{,\varphi},
            \label{ORiemann}\\
	&\overline{R}_{\mu\nu} 
	= R_{\mu\nu}
		- \omega^{-1}\bigl[g_{\mu\nu} g^{\rho\sigma} 
            + (d-2)\delta_\mu^\rho\delta_\nu^\sigma\bigr]
		\omega_{;\rho\sigma}
        \nonumber\\ &\qquad\quad
		+ \omega^{-2}\bigl[
            2(d-2)\delta_\mu^\rho\delta_\nu^\sigma
            - (d-3) g_{\mu\nu} g^{\rho\sigma}
        \bigr] \omega_{,\rho} \omega_{,\sigma},
        \label{ORicci}\\
	&\overline{R} 
	= \omega^{-2} R 
		-  2 (d-1) \omega^{-3}\  \omega^{; \mu}{}_\mu
		-  (d-4) (d-1)\omega^{-4}\ \omega_{,\mu} \omega^{,\mu}.
        \label{OCurv}
\end{align}
In the above equations the semi-colon refers to the covariant derivation 
    with respect to $g$, $\Gamma$ is the Levi-Civita connection, and
    $[\alpha\beta] = \alpha\beta - \beta\alpha$ 
    is the antisymmetric part of the enclosed indices.

An equation, depending on the metric and symbolically written, $E(A)=0$
is said to be Weyl invariant if and only if there exists a conformal 
weight $h(A)\in\mathbb{R}$ such that
\begin{equation*}
    \overline{E}(\overline{A}) = \omega^{h'} E(A), \qquad 
    \overline{A} = \omega^h A.
\end{equation*}
That is, for a given solution $A$ of the equation $E(A) = 0$ the rescaled 
    field $\overline{A} = \omega^h A$ is a solution of 
    $\overline{E}(\overline{A})=0$ on the Weyl related spacetime 
    $(M,\overline{g})$.

\section{The residual gauge freedom}
\label{GFEqEff}

Equation \eqref{EQFG} provides a conformally invariant gauge fixing equation of
Eq.~\eqref{EqPrinc}.
Here it is shown that $\varphi$ in Eq.~\eqref{GaugeInv} is no longer arbitrary,
its remaining gauge freedom is found and commented upon.

\subsection{The equation fulfilled by the gauge scalar $\varphi$}

To begin let us consider a pure gauge field
\begin{equation}
    \label{Fgauge}
    A = \STG^s( \varphi ).
\end{equation}
Then, plugging Eq.~\eqref{Fgauge} into Eq.~\eqref{EQFG} yields
\begin{align}
    \Bigl(\square - \frac{1}{6} R\Bigr) \nabla^s \STG^s (\varphi)
    &= \Bigl(\square - \frac{1}{6} R\Bigr) \nabla^{s-1} (\nabla\cdot\STG)
        (\STG^{s-1}(\varphi )) \label{constrain.1} \\
    &= \Bigl(\square - \frac{1}{6} R\Bigr) \nabla^{s-1} U^{s-1}
        (\STG^{s-1}(\varphi )) \label{constrain.2}\\
    &= \alpha_s \Bigl(\square - \frac{1}{6} R\Bigr) U^{s-1}{}' 
        (\nabla^{s-1}\STG^{s-1}(\varphi )) \label{constrain.3}
\end{align}
with 
\begin{equation}
    \label{Up}
    U^{s} (A) = \square A 
    + \Bigl(\frac{d+2s-4}{d+2s-2}\Bigr) \STG ( \nabla\cdot A )
    + \frac{s(d+s-2)}{d(d-1)} R A
\end{equation}
for $A$ symmetric traceless of rank $s$ and for $\phi$ a scalar field
\begin{equation}
    \label{Upp}
    U^{s}{}'(\phi)
     = \Bigl(\square + \frac{s(d-1 +s)}{d(d-1)} R\Bigr) \phi,
\end{equation}
and $\alpha_s$ a non-vanishing numerical factor, namely 
$\alpha_s = (1+s)(d+s-2)(d+2s-2)^{-1}$.
Going from Eq.~\eqref{constrain.2} to Eq.~\eqref{constrain.3}, that is obtaining
Eq.~\eqref{Upp} from Eq.~\eqref{Up}, is performed in the same vein as the
computation already carried in Sec.~\ref{SubSecDerivGFE}, for which
the commutation relations between divergences and a second order equation akin
to Eq.~\eqref{Up} were obtained.

Then, carrying this scheme in Eq.~\eqref{constrain.1} till its end with
Eq.~\eqref{Upp} and minding that $d=4$, one gets that the scalar gauge field
$\varphi$ fulfills
\begin{equation}
	\label{eq.GJMS1}
	\Bigl(\square - \frac{1}{6} R\Bigr) 
		\square 
		\Bigl( \square + \frac{1}{3} R\Bigr)
		\times\dots 
		\times \Bigl( \square + \frac{(s-1)(s+2)}{12} R\Bigr) \varphi = 0.
\end{equation}
This equation is known as \emph{Branson's factorization formula\cite{Branson1995, Graham2007} of GJMS\cite{GJMS1, GJMS2, GJMS3} operators} (on CFES) and is conformally invariant.

\subsection{
    The restricted invariance of Eq.~\eqref{eq.GJMS1} and of pure gauge solutions}

This subsection provides a direct demonstration of the conformal invariance of Eq.~\eqref{eq.GJMS1}, emulating the calculus used for Eq.~\eqref{eqPaneitz} via Eq.~\eqref{GaugeEastwoodSinger}. 
Importantly, this computation hinges completely{\em} on the accidental occurrence of both conformal and gauge invariance for $d=4$.

\begin{property}
    Equation~\eqref{eq.GJMS1} is invariant under restricted Weyl transformations, 
    the space of pure gauge solutions to Eq.~\eqref{EqPrinc} remains invariant under the action of SO$_0(2,4)$.
\end{property}

\begin{proof}
The main idea of the proof is that the gauge fixing condition \eqref{EQFG} is invariant under restricted Weyl transformations on the space of solutions of Eq.~\eqref{EqPrinc} for $d=4$ (cf. Prop.~\ref{prop:Restricted_GF}).
Then, if one plugs in Eq.~\eqref{EQFG} a pure gauge solution the 
\emph{``on the space of solutions of Eq.~\eqref{EqPrinc}''} part is taken care of and one is left to look if there is a Weyl rescaling of the
scalar field $\varphi$ resulting (only) in the appropriate rescaling of the field, that is:
with $h(A) = 1-s-d/2$, does there exist $w\in\mathbb{R}$ such that
\begin{equation}
    \label{QuestionGene}
    \overline{\STG^s}(\overline{\varphi})
        = \overline{\STG^s}(\omega^{w}\varphi)
        = \omega^{h} \STG^s(\varphi)\ ?
\end{equation}
If so the restricted conformal invariance is proven.

Setting $\rho = \omega^{-1}$, $v = - w$ and noticing the identity:
\begin{equation*}
    \overline{\STG}(f) = \rho^2 \STG(f),
\end{equation*}
simplifies the right hand side of Eq.~\eqref{QuestionGene} which might then be
expressed as
\begin{align*}
    \overline{\STG^s}(\omega^{w}\varphi) 
    &= \overline{\STG^s}(\rho^{v}\varphi)
    = \underbrace{\rho^2\STG(\rho^2\STG(\dots }_{s\text{ times}}
    (\rho^{v}\varphi )\dots )).
\end{align*}
To get a tractable formula let us use the following notation
\begin{align*}
    \STt\ :\ f_1, f_2 &\mapsto \frac{(s_1 + s_2)!}{s_1! s_2!} (f_1\cdot f_2)\Big\rvert_{ST}
\end{align*}
in which $\rvert_{ST}$ is the symmetric traceless projector (i.e. 
$\rvert_{ST}\rvert_{ST}=\rvert_{ST}$).
On scalars the product $\STt$ reduces to the pointwise product of two
functions, as is also the case for the product between a scalar and a vector.
Then, thanks to this definition, accommodated to that of the symmetric traceless
gradient (STG) given in Eq.~\eqref{defSTG}, one has a Leibniz identity
\begin{equation*}
    \STG(f \STt g) = \STG(f)\STt g + f\STt\STG(g).
\end{equation*}
Then, beginning with $\rho^{v}\varphi = \rho^{v}\STt\varphi$, using
Eq.~\eqref{WeylRestr} to discard any term such as $\STG^i(\rho)$ with
$i\geqslant 2$ and an induction on the degree yields
\begin{equation*}
    \overline{\STG^s}(\rho^{v}\varphi)
    = \sum_{i=0}^s \frac{\Gamma(v+s)}{\Gamma(v+s-i)} \binom{s}{i} 
        \rho^{v + 2s-i}
        (\STG(\rho))^{\STt i}\ \STt \STG^{s-i}(\varphi).
\end{equation*}
Finally, setting $w = s-1$, that is $v = 1-s$, to cancel the terms with
$i\geqslant 1$, and restoring the $\omega$'s provides the identity
\begin{equation}
    \label{ReponseGene}
    \overline{\STG^s}(\omega^{s-1}\varphi) = \omega^{-s-1}\STG^s(\varphi).
\end{equation}

However, for Eq.~\eqref{eq.GJMS1} to be conformally
invariant the scaling factor of the resulting field in Eq.~\eqref{ReponseGene}
has to agree with  that of $A$, namely:
    $h(A) = 1 - s - (d/2) = - s - 1$,
which is achieved only for $d=4$.

The SO$_0(2,4)$ invariance of Eq.~\eqref{eq.GJMS1} is then inherited from that
of $\square^n$ on Minkowski spacetime, a well known result.\cite{JakobsenVergne}
\end{proof}

\begin{remark}
Equation~\eqref{ReponseGene} can be
interpreted in the following manner: a pure gauge solution on $(M,g)$ is
mapped on a pure gauge solution on $(M,\overline{g})$.
\end{remark}

\subsection{Remark on the residual gauge freedom on de Sitter}

If the underlying CFES is the de Sitter spacetime the equations
governing the residual gauge invariance are linked to peculiar
scalar representations of the de Sitter group SO$_0(1,d)$. 
Indeed, in such a case the first order Casimir operator $\mathcal{Q}_1$ in the
scalar representation of SO$_0(1,d)$ is related to the Laplace-Beltrami operator
through $\mathcal{Q}_1 = d(d-1)R^{-1} \square$.
Then, Eq.~\eqref{Upp} can be recast as
\begin{equation*}
    [\mathcal{Q}_1 + j(d-1+j)] \phi = 0, 
	\qquad j\in\mathbb{N}.
\end{equation*}
This, precisely, is the equation fulfilled by a scalar field in the discrete 
series of SO$_0(1,d)$.
As a consequence, if the gauge scalar $\varphi$ transforms covariantly under
the de Sitter group, it decomposes as
\begin{equation*}
	\varphi = \varphi_{\text{cc}} \oplus
    \biggl(\bigoplus_{j=0}^{s-1} \varphi_{\text{ds} (j)}\biggr),
\end{equation*}
in which $\varphi_{\text{cc}}$ stands for the massless conformally coupled
field,\cite{Huguet:2006fe} which lies in the complementary series of
SO$_0(1,d)$, and $\varphi_{\text{ds} (j)}$ is the $j$th term in the discrete
series of SO$_0(1,d)$.\cite{Dixmier,Takahashi}
The $0$th term of the discrete series is the infamous massless minimally
coupled field 
\cite{Allen1985,Allen1987,Kirsten:1993ug,Gazeau:1999mi,Joung:2007je} and fields
further in the discrete series are the so-called scalar tachyons.\cite{Folacci:1992xc,Folacci:1996dv,Bros:2010wa}

\section{The special case of $d=2$}
\label{Sec2D}

Dimension two is far too peculiar with respect to conformal invariance
and cannot be examined on the same footing as the others.
This section describes the lack of content of Eq.~\eqref{EqPrinc} for $d=2$.

One could consider Eq.~\eqref{EqPrinc} and wonder what happens for $d=2$.
First, the restricted conformal invariance remains. 
However, following the same steps as in the main part of the article, one
would find that the equation admits solutions determined up to a vector (as then
$(1+a_2)\rvert_{d=2} = 0$).
Similarly, one could get the impression that the set
\begin{equation*}
	\left\{
    \begin{aligned}
		E_s(A) &= 0, \\
		E_1(V) &= 0,
    \end{aligned}\right.
	\qquad 
	(d=2)
\end{equation*}
where $V^{\mu_1} = \nabla_{\mu_2}\dots\nabla_{\mu_s} A^{\mu_1\dots\mu_s}$, is
conformally invariant while restricting the (new-found) gauge freedom.
While the former is true, the latter is \emph{not}.

In order to show this, note first that $A$ possesses only two independent
components, whatever the rank $s$.
In the usual Cartesian coordinates let us choose those as
\begin{equation*}
	\Phi^{\pm} = A^{00\dots 00} \pm A^{00\dots 01}.
\end{equation*}
Then, turning to the chiral coordinates $x^{\pm} = x^0 \pm x^1$
Eq.~\eqref{EqPrinc} is rewritten as
\begin{equation}
    \label{eq.2D}
    \left\{
    \begin{aligned}
        \partial_{+}^2 \Phi^ {+} &= 0, \\
        \partial_{-}^2 \Phi^ {-} &= 0.
    \end{aligned}\right.
\end{equation}
The system is conformally invariant and the rank from which $\Phi^\pm$ emanates
is read off the way the field transforms.
The field admits the gauge transformation
\begin{equation*}
	A \mapsto {}^aA = A + \STG^{s-1}(a), \quad s \geqslant 2,
\end{equation*}
with $a$ an arbitrary vector field.
On the independent components $\Phi^\pm$ the equation above reads as
\begin{equation}
	\label{eq.transfgauge2d}
	{}^a \Phi^{\pm} = \Phi^{\pm}  + s!\, \partial_{\mp}^{s-1} a^{\pm}, \quad
	a^{\pm} = a^0 \pm a^1.
\end{equation}
Now, for a given field $ A \simeq (\Phi^{+}, \Phi^{-})$ one can find $(a^{+},
a^{-})$ such that the gauge transformed field \eqref{eq.transfgauge2d} is null.
This means that, for $s\geq 2$, any field $A$ is in the (gauge) equivalence
class of the trivial solution.
Hence that field carries no physical content.

\section{Proofs}
\label{sec:Proofs}

\subsection{Proof of the restricted Weyl invariance of the gauge fixing equation (Prop.~\ref{prop:Restricted_GF})}
\label{app:Restricted_GF}

\begin{proof}
Let us consider first
\begin{equation*}
    \overline{\phi}
    = \overline{\nabla}_{\mu_1}\dots\overline{\nabla}_{\mu_s}
        \overline{A}^{\mu_1\dots\mu_s}
    = \overline{\nabla}_{\mu_1}\dots\overline{\nabla}_{\mu_s}
        \omega^h A^{\mu_1\dots\mu_s},
\end{equation*}
in which $h(A^{\mu_1\dots\mu_s}) = 1 - s - d/2$.
Equation \eqref{WeylRestr} can be rewritten as
\begin{equation}
    \label{WeylRestr2}
        \Bigl(\nabla_\mu\nabla_\nu - \frac{1}{d}g_{\mu\nu} \square\Bigr) \rho 
        = -\omega^{-2}\Bigl(\overline{\nabla}_\mu\overline{\nabla}_\nu 
            - \frac{1}{d}\overline{g}_{\mu\nu} \overline{\square}\Bigr) \omega
        = 0,
\end{equation}
with $\rho = 1/\omega$.
Thanks to the tracelessness of $A$ one realizes that there cannot be
derivatives of $\omega$ of degree greater or equal to 2 contracted with $A$ 
since
\begin{equation*}
    (\overline{\nabla}_{\mu_i}\overline{\nabla}_{\mu_j} \omega )
        A^{\mu_1\dots\mu_s}
    = \left[\left(
        \overline{\nabla}_{\mu_i}\overline{\nabla}_{\mu_j} 
        - \frac{1}{d}\overline{g}_{\mu_i\mu_j} \overline{\square}
      \right)\omega\right] A^{\mu_1\dots\mu_s}
    = 0,
\end{equation*}
in which one uses the tracelessness of $A$ and then that Eq.~\eqref{WeylRestr2}
is fulfilled by $\omega$.
This simplifies greatly the expansion of $\overline{\phi}$ as one then gets
\begin{align}
    \overline{\phi}
    &= \sum_{i=0}^s \frac{\Gamma(h+1)}{\Gamma(h+1-i)} \binom{s}{i}
        \omega^{h-i}
        (\overline{\nabla}\omega )^i \overline{\nabla}{}^{s-i} A\nonumber\\
    &= \sum_{i=0}^s \frac{\Gamma(h+1)}{\Gamma(h+1-i)} \binom{s}{i}
        \omega^{h-i}
        (\nabla\omega )^i \overline{\nabla}{}^{s-i} A,\label{PhiFp2}
\end{align}
using the notation in which an index contracted with $A$ is not written.
For instance a generic term in Eq.~\eqref{PhiFp2} reads as
\begin{equation}
    \label{notationsansindices}
    (\nabla\omega )^i \overline{\nabla}{}^{s-i} A
    = (\nabla_{\mu_1} \omega) \dots  (\nabla_{\mu_i} \omega) 
        (\overline{\nabla}{}_{\mu_{i+1}}\dots \overline{\nabla}{}_{\mu_s}
        A^{\mu_1\dots\mu_s}).
\end{equation}
Now, one can express $\overline{\nabla}{}^n A$ in terms of $\nabla$ and $\omega$.
By induction on $n$ one gets
\begin{align}
    \overline{\nabla}{}^n A 
    &= \sum_{i=0}^n \frac{\Gamma(d+2s-n-1+i)}{\Gamma(d+2s-n-1)}\binom{n}{i}
        (-1)^i \rho^{-i }(\nabla\rho)^i \nabla^{n-i} A \nonumber \\
    &= \sum_{i=0}^n \frac{\Gamma(d+2s-n-1+i)}{\Gamma(d+2s-n-1)}\binom{n}{i}
        \omega^{-i }(\nabla\omega)^i \nabla^{n-i} A, \label{PhiFp3}
\end{align}
using the same argument about the derivatives of $\rho$, with
Eq.~\eqref{WeylRestr2}, and finally
$\omega^{-1}(\nabla\omega) = -\rho^{-1}(\nabla\rho)$ to recast the result in
terms of $\omega$ solely.
Then, plugging Eq.~\eqref{PhiFp3} in Eq.~\eqref{PhiFp2} and inverting the order
of summation yields
\begin{align}
    \overline{\phi} 
    &= \sum_{i=0}^s \left[\sum_{j=0}^i  \binom{i}{j} 
        \frac{\Gamma(h+1)}{\Gamma(h+1-j)}\frac{\Gamma(d+s-1+i)}{\Gamma(d+s-1+j)}
        \right]
        \binom{s}{i} \omega^{h-j} (\nabla\omega )^i \nabla^{s-i} A
        \nonumber\\
    &= \sum_{i=0}^s \frac{\Gamma(d+s-1+i)}{\Gamma(d+s-1)}
        {}_2F_1(-i, -h; d+s-1;1)
        \binom{s}{i} \omega^{h-j} (\nabla\omega )^i \nabla^{s-i} A
        \nonumber\\
    &= \sum_{i=0}^s \frac{s!}{s-i!}(i+1)\ \omega^{-1-s-i}
        (\nabla\omega )^i \nabla^{s-i} A, \label{PhiFp5}
\end{align}
in which $h$ has been set to its value,  $d=4$ and  ${}_2F_1$ is a hypergeometric
function.

Replacing $\overline{\phi}$ by Eq.~\eqref{PhiFp5} and applying the conformal
laplacian to it yields
\begin{align*}
    \Bigl(\overline{\square} -\frac{1}{6}\overline{R}\Bigr) \overline{\phi}
    &= \Bigl(\overline{\square} -\frac{1}{6}\overline{R}\Bigr)
        \Bigl(\sum_{i=0}^s \frac{s!}{s-i!}(i+1)\ \omega^{-1-s-i}
        (\nabla\omega )^i \nabla^{s-i} A\Bigr)\\
    &= \sum_{i=0}^s \frac{s!}{s-i!}(i+1) \omega^{-3}
            \Bigl(\square -\frac{1}{6}R\Bigr)
            \omega^{-s-i}(\nabla\omega)^i \nabla^{s-i} A,
\end{align*}
using the simplying fact that $(\overline{\square} - \overline{R}/6)\omega^{-1}f
    = \omega^{-3}(\square - R/6)f$, for $d=4$.
Now, brute-forcing the right-hand side through the differential operator yields
\begin{align*}
    \Bigl(\overline{\square} -\frac{1}{6}\overline{R}\Bigr) \overline{\phi}
    = \sum_{i=0}^s \frac{s!}{s-i!}&(i+1) \omega^{-s-i-3}\Bigl\{
    (\nabla\omega)^i\Bigl(\square -\frac{1}{6}R\Bigr)\nabla^{s-i} A\\
    &-2(s+i)\omega^{-1}(\N_\alpha\omega)(\N\omega)^i\N^\alpha\N^{s-i}A\\
    &+(s+i)(s+i+1)\omega^{-2}(\N_\alpha\omega)(\N^\alpha\omega)(\nabla\omega)^i\nabla^{s-i}A\\
    &-(s+i)\omega^{-1}(\square\omega)(\N\omega)^i\N^{s-i}A\\
    &-2i(s+i)\omega^{-1}(\N\omega)^{i-1}(\N_\alpha\N\omega)(\N^\alpha\omega)\N^{s-i}A\\
    &+i(i-1)(\N\omega)^{i-2}(\N_\alpha\N\omega)(\N^\alpha\N\omega)\N^{s-i}A\\
    &+2i(\N\omega)^{i-1}(\N_\alpha\N\omega)\N^\alpha\N^{s-i}A\\
    &+i(\N\omega)^{i-1}(\square\N\omega)\N^{s-i}A\Bigr\},
\end{align*}
which is quite unappealing.
However, using Eq.~\eqref{WeylRestr} and Eq.~\eqref{WeylRestr1}  as
\begin{align*}
    &(\N_\alpha\N\omega)
    = 2\omega^{-1}(\N_\alpha\omega)(\N\omega)
    - \frac{1}{2} g_{\cdot\alpha}(\N_\beta\omega)(\N^\beta\omega)
    + \frac{1}{4} g_{\cdot\alpha}(\square\omega),\\
    &(\square\N\omega)
    = 3\omega^{-1}(\N\omega)(\square\omega)
    - \frac{R}{12}(\N\omega),
\end{align*}
and $(\N^2\omega) = 2\omega^{-1}(\N\omega)^2$,
in the last four lines recasts the equation as
\begin{align*}
    \Bigl(\overline{\square} -\frac{1}{6}\overline{R}\Bigr) \overline{\phi}
        = \sum_{i=0}^s \frac{s!}{s-i!}&(i+1) \omega^{-s-i-3}(\nabla\omega)^i\Bigl\{
        \Bigl(\square -\frac{2+i}{12}R\Bigr)\nabla^{s-i} A\\
    &-2(s-i)\omega^{-1}(\N_\alpha\omega)\N^\alpha\N^{s-i}A\\
    &+(s-i)(s-i-1)\omega^{-2}(\N_\alpha\omega)(\N^\alpha\omega)\nabla^{s-i}A
    \Bigr\},
\end{align*}
shifting the summation variable in the last two lines lets us recognize, abusing a bit the notation, the contraction of $(\nabla\omega)^i$ with $\STG\nabla\ \nabla^{s-i}A$, thus giving
\begin{align*}
    \Bigl(\overline{\square} -\frac{1}{6}\overline{R}\Bigr) \overline{\phi}
    &
    = \sum_{i=0}^s \frac{s!}{s-i!}(i+1) \omega^{-s-i-3} (\nabla\omega)^i
    \Bigl( \square - \frac{2}{i+1}\STG\nabla -\frac{2+i}{12} R\Bigl) 
    \nabla^{s-i} A \\
    &= \sum_{i=0}^s \frac{s!}{s-i!}(i+1) \omega^{-s-i-3} (\nabla\omega)^i
           E_i( \nabla^{s-i} A).
\end{align*}
On the space of solutions of $E_s(A) = 0$, according to Eqs.~\eqref{ID.Div1} and \eqref{ID.Div2},
each term with $1\leqslant i \leqslant s$ vanishes thus leaving
\begin{equation*}
    \Bigl(\overline{\square} -\frac{1}{6}\overline{R}\Bigr) \overline{\phi} 
    = \omega^{-s-3} \Bigl(\square -\frac{1}{6}R\Bigr)\phi.
    \qedhere
\end{equation*}
\end{proof}

\subsection{Proof of the invertibility of Eqs.~\eqref{def.MN} and \eqref{def.EB} (Prop.~\ref{prop:InversionEB})}
\label{InversionFEB}

\begin{proof}
This is a two-steps proof in which two changes of basis are successively inverted. 

\noindent\emph{1. The $M_j$-$N_j$'s in terms of the $E_j$-$B_j$'s.}
From the properties fulfilled by $F$, recorded in Eq.~\eqref{ID.FS}, and the
definition
\eqref{def.MN} one finds that
\begin{subequations}
    \label{recMN}
    \begin{align}
        \delta_{i_ai_b} M_j^{i_1\dots i_j}
            &= M_{j-2}^{i_1\dots \widehat{i_a}\dots \widehat{i_b}\dots i_j},\\
        \delta_{i_ai_b} N_j^{i_1\dots i_j}
            &= N_{j-2}^{i_1\dots \widehat{i_a}\dots \widehat{i_b}\dots i_j},
    \end{align}
\end{subequations}
in which a hat over an index means that said index is omitted.
From Eqs.~\eqref{def.EB} and \eqref{recMN} one can invert the $M_j$'s in
terms of the $E_j$'s since 
\begin{equation}
    \label{eq.ETM}
    E_j^{i_1\dots i_j}
    = \sum_{k = 0}^{\lfloor \frac{j}{2}\rfloor} \alpha_{jk}
        \delta^{(i_1 i_2}\dots\delta^{i_{2k-1}i_{2k}}
        M_{j-2k}^{i_{2k+1}\dots i_j)},
\end{equation}
in which the coefficients $\alpha_{jk}$ are determined by asking that the right
hand side of Eq.~\eqref{eq.ETM} is traceless, which is well posed thanks to
Eq.~\eqref{recMN} once one has chosen an $\alpha_{j0}$ (here $\alpha_{j0} = 1$
for $E$ and $M$ and $\alpha_{j0} = 1/(s+1)$ for $B$ and $N$).
  
Then, Eq.~\eqref{eq.ETM} might be viewed as an upper triangular
transformation between the $E_j$'s and the $M_j$'s and as such can be inverted
as
\begin{align*}
    M_j^{i_1\dots i_j} 
    &= E_j^{i_1\dots i_j} 
    - \alpha_{j1} \delta^{(i_1i_2}_{} E_{j-2}^{i_3\dots i_j)}
    - (\alpha_{j2} - \alpha_{j1}\alpha_{j-2\ 1})
        \delta^{(i_1i_2}\delta^{i_3i_4}E_{j-4}^{i_5\dots i_j)} - \dots\\
    &= \sum_{k=0}^{\lfloor \frac{j}{2}\rfloor} \beta_{jk} 
        \delta^{(i_1 i_2}\dots\delta^{i_{2k-1}i_{2k}}
        E_{j-2k}^{i_{2k+1}\dots i_j)}.
\end{align*}
The same arguments apply to the inversion of the $B_j$'s in terms of the
$N_j$'s.
With a bit of perseverance one finds that
\begin{align*}
    \alpha_{jk} 
        &= (-1)^k \frac{j!}{j-2k!}
            \frac{2j-1-2k!!\ 2k-1!!}{2j -1 !!}
        = (-1)^k \frac{j!\ j!\ 2j-2k!\ 2k!}
            {2j!\ j-k!\ j-2k!\ k!}, \\ 
    \beta_{jk}
        &=  \frac{j!}{j-2k!}
            \frac{2j-4k +1!!\ 2k -1!!}{2j - 2k +1 !!} 
         = \frac{j!\ 2j - 4k + 2!\ 2k!\ j - k + 1!}
            {k!\ j - 2k!\ j - 2k + 1!\ 2j - 2k + 2!},
\end{align*}
for $k\geqslant 1$ and $\alpha_{j0} = \beta_{j0} = 1$ otherwise.
In the above we used the identity on bifactorials of odd numbers:
$(2n - 1)!! = 2n! / 2^n n!$.
This completes the first step.

\noindent\emph{2. $F$ in terms of the $M_j$-$N_j$'s.}
We are now concerned with the second step of the inversion, that is to write an
arbitrary component of $F$ in terms of the $M_j$'s and the $N_j$'s.
From the definition \eqref{def.MN} an arbitrary $F^{0\ 0\dots 0i_1\dots i_j}$
might already be written in terms of the $M_j$'s.
What is left to show is that $F^{l\ 0\dots 0i_1\dots i_j}$ might also be written
in terms of the $M_j$'s and the $N_j$'s.

From the symmetries of $F$, cf. Eq.~\eqref{ID.FS}, one can write the following
identity:
\begin{equation}
    \label{subidFS}
    F^{l\ i_1\dots i_j 0\dots 0} 
    = \frac{j}{j+1} F^{[l\ (i_1]  i_2\dots i_j) 0\dots 0}
    - \Bigl(\frac{s-j}{j+1}\Bigr) F^{0\  0\dots 0 i_1\dots i_j l},
\end{equation}
in which the second term of the right hand side, from the above remark, is already
inverted in terms of the $M_j$'s.
Then, it suffices to show that one can express the first term unambiguously in terms of the $M_j$'s and the $N_j$'s.

Out of the definition \eqref{def.MN} of the $N_j$'s, of the properties fulfilled
by the antisymmetric tensor $\varepsilon_{ijk}$ and those of $F$ [given in 
Eq.~\eqref{ID.FS}] we obtain the following identity:
\begin{multline*}
    \varepsilon_{}^{kl(i_1}N_j^{i_2\dots i_j) k}
    = j F^{[l\ (i_1] i_2\dots i_j) 0\dots 0}
    - (s+1)\delta^{l(i_1} F^{\lvert 0\rvert\ i_2\dots i_j) 0\dots 0} \\
    - (j-2)\delta^{(i_1i_2} F^{[\lvert l\rvert\ i_3] i_4\dots i_j)0\dots 0}
    + (s+1)\delta^{(i_1i_2} F^{\lvert 0\rvert\ i_3\dots i_j)l 0\dots 0}
\end{multline*}
put in a more suitable form for our purpose as
\begin{multline*}
    j F^{[l\ (i_1] i_2\dots i_j) 0\dots 0}
    = (j-2)\delta^{(i_1i_2} F^{[\lvert l\rvert\ i_3] i_4\dots i_j)0\dots 0}
    + \varepsilon^{kl(i_1}N_j^{i_2\dots i_j) k}\\
    +(s+1)(\delta^{l(i_1} M_{j-1}^{i_2\dots i_j)} 
        -\delta^{(i_1i_2} M_{j-1}^{i_3\dots i_j)l}).
\end{multline*}
By recursively using the above formula one can invert $F^{[l\ (i_1] \dots}$ in
terms of the $M_j$'s and the $N_j$'s terminating with either
\begin{equation*}
    F^{[l\,i_1]0\dots 0} = \varepsilon^{kli_1} N_1^{k}
\end{equation*}
if $j$ is odd, or
\begin{multline*}
    F^{[l\ i_1] i_2 0\dots 0} + F^{[l\ i_2] i_10\dots 0}
    = \frac{1}{2}(\varepsilon^{kli_1} N_2^{i_2k}
    + \varepsilon^{kli_2} N_2^{i_1k})
    + \frac{1}{2}(s+1)(\delta^{li_1}M_1^{i_2}
    + \delta^{li_2}M_1^{i_1}-2\delta^{i_1i_2} M_1^l)
\end{multline*}
if $j$ is even.

This means that the first term in Eq.~\eqref{subidFS} can unambiguously be
written in terms of the $N_j$'s and the $M_j$'s.
From Eq.~\eqref{def.MN} it is also the case for the second term in
Eq.~\eqref{subidFS}.
Hence $F^{l\ 0\ldots 0i_1\ldots i_j}$ has a well posed decomposition over the
$M_j$'s and the $N_j$'s.
Finally, $F$ can be written in terms of the $M_j$-$N_j$'s, which
themselves can be written in terms of the $E_j$-$B_j$'s.
This completes the second step.
\end{proof}

\bibliography{references}

\begin{thebibliography}{38}%
\makeatletter
\providecommand \@ifxundefined [1]{%
 \@ifx{#1\undefined}
}%
\providecommand \@ifnum [1]{%
 \ifnum #1\expandafter \@firstoftwo
 \else \expandafter \@secondoftwo
 \fi
}%
\providecommand \@ifx [1]{%
 \ifx #1\expandafter \@firstoftwo
 \else \expandafter \@secondoftwo
 \fi
}%
\providecommand \natexlab [1]{#1}%
\providecommand \enquote  [1]{``#1''}%
\providecommand \bibnamefont  [1]{#1}%
\providecommand \bibfnamefont [1]{#1}%
\providecommand \citenamefont [1]{#1}%
\providecommand \href@noop [0]{\@secondoftwo}%
\providecommand \href [0]{\begingroup \@sanitize@url \@href}%
\providecommand \@href[1]{\@@startlink{#1}\@@href}%
\providecommand \@@href[1]{\endgroup#1\@@endlink}%
\providecommand \@sanitize@url [0]{\catcode `\\12\catcode `\$12\catcode
  `\&12\catcode `\#12\catcode `\^12\catcode `\_12\catcode `\%12\relax}%
\providecommand \@@startlink[1]{}%
\providecommand \@@endlink[0]{}%
\providecommand \url  [0]{\begingroup\@sanitize@url \@url }%
\providecommand \@url [1]{\endgroup\@href {#1}{\urlprefix }}%
\providecommand \urlprefix  [0]{URL }%
\providecommand \Eprint [0]{\href }%
\providecommand \doibase [0]{https://doi.org/}%
\providecommand \selectlanguage [0]{\@gobble}%
\providecommand \bibinfo  [0]{\@secondoftwo}%
\providecommand \bibfield  [0]{\@secondoftwo}%
\providecommand \translation [1]{[#1]}%
\providecommand \BibitemOpen [0]{}%
\providecommand \bibitemStop [0]{}%
\providecommand \bibitemNoStop [0]{.\EOS\space}%
\providecommand \EOS [0]{\spacefactor3000\relax}%
\providecommand \BibitemShut  [1]{\csname bibitem#1\endcsname}%
\let\auto@bib@innerbib\@empty
\bibitem [{\citenamefont {Deser}\ and\ \citenamefont
  {Nepomechie}(1984)}]{Deser:1983mm}%
  \BibitemOpen
  \bibfield  {author} {\bibinfo {author} {\bibfnamefont {S.}~\bibnamefont
  {Deser}}\ and\ \bibinfo {author} {\bibfnamefont {R.~I.}\ \bibnamefont
  {Nepomechie}},\ }\bibfield  {title} {\enquote {\bibinfo {title} {{Gauge
  Invariance Versus Masslessness in De Sitter Space}},}\ }\href
  {https://doi.org/10.1016/0003-4916(84)90156-8} {\bibfield  {journal}
  {\bibinfo  {journal} {Annals Phys.}\ }\textbf {\bibinfo {volume} {154}},\
  \bibinfo {pages} {396} (\bibinfo {year} {1984})}\BibitemShut {NoStop}%
\bibitem [{\citenamefont {Binegar}, \citenamefont {Fronsdal},\ and\
  \citenamefont {Heidenreich}(1983{\natexlab{a}})}]{Binegar1982Mink}%
  \BibitemOpen
  \bibfield  {author} {\bibinfo {author} {\bibfnamefont {B.}~\bibnamefont
  {Binegar}}, \bibinfo {author} {\bibfnamefont {C.}~\bibnamefont {Fronsdal}},\
  and\ \bibinfo {author} {\bibfnamefont {W.}~\bibnamefont {Heidenreich}},\
  }\bibfield  {title} {\enquote {\bibinfo {title} {{Conformal QED}},}\ }\href
  {https://doi.org/10.1063/1.525664} {\bibfield  {journal} {\bibinfo  {journal}
  {J. Math. Phys.}\ }\textbf {\bibinfo {volume} {24}},\ \bibinfo {pages} {2828}
  (\bibinfo {year} {1983}{\natexlab{a}})}\BibitemShut {NoStop}%
\bibitem [{\citenamefont {Binegar}, \citenamefont {Fronsdal},\ and\
  \citenamefont {Heidenreich}(1983{\natexlab{b}})}]{Binegar1982AdS}%
  \BibitemOpen
  \bibfield  {author} {\bibinfo {author} {\bibfnamefont {B.}~\bibnamefont
  {Binegar}}, \bibinfo {author} {\bibfnamefont {C.}~\bibnamefont {Fronsdal}},\
  and\ \bibinfo {author} {\bibfnamefont {W.}~\bibnamefont {Heidenreich}},\
  }\bibfield  {title} {\enquote {\bibinfo {title} {{de Sitter QED}},}\ }\href
  {https://doi.org/10.1016/0003-4916(83)90197-5} {\bibfield  {journal}
  {\bibinfo  {journal} {Ann. Phys.}\ }\textbf {\bibinfo {volume} {149}},\
  \bibinfo {pages} {254} (\bibinfo {year} {1983}{\natexlab{b}})}\BibitemShut
  {NoStop}%
\bibitem [{\citenamefont {Bayen}\ \emph {et~al.}(1985)\citenamefont {Bayen},
  \citenamefont {Flato}, \citenamefont {Fronsdal},\ and\ \citenamefont
  {Haidari}}]{Bayen:1984dt}%
  \BibitemOpen
  \bibfield  {author} {\bibinfo {author} {\bibfnamefont {F.}~\bibnamefont
  {Bayen}}, \bibinfo {author} {\bibfnamefont {M.}~\bibnamefont {Flato}},
  \bibinfo {author} {\bibfnamefont {C.}~\bibnamefont {Fronsdal}},\ and\
  \bibinfo {author} {\bibfnamefont {A.}~\bibnamefont {Haidari}},\ }\bibfield
  {title} {\enquote {\bibinfo {title} {{Conformal Invariance and Gauge Fixing
  in {QED}}},}\ }\href {https://doi.org/10.1103/PhysRevD.32.2673} {\bibfield
  {journal} {\bibinfo  {journal} {Phys. Rev.}\ }\textbf {\bibinfo {volume}
  {D32}},\ \bibinfo {pages} {2673} (\bibinfo {year} {1985})}\BibitemShut
  {NoStop}%
\bibitem [{\citenamefont {Faci}\ \emph {et~al.}(2009)\citenamefont {Faci},
  \citenamefont {Huguet}, \citenamefont {Queva},\ and\ \citenamefont
  {Renaud}}]{Faci2009}%
  \BibitemOpen
  \bibfield  {author} {\bibinfo {author} {\bibfnamefont {S.}~\bibnamefont
  {Faci}}, \bibinfo {author} {\bibfnamefont {E.}~\bibnamefont {Huguet}},
  \bibinfo {author} {\bibfnamefont {J.}~\bibnamefont {Queva}},\ and\ \bibinfo
  {author} {\bibfnamefont {J.}~\bibnamefont {Renaud}},\ }\bibfield  {title}
  {\enquote {\bibinfo {title} {{Conformally covariant quantization of Maxwell
  field in de Sitter space}},}\ }\href
  {https://doi.org/10.1103/PhysRevD.80.124005} {\bibfield  {journal} {\bibinfo
  {journal} {Phys. Rev.}\ }\textbf {\bibinfo {volume} {D80}},\ \bibinfo {pages}
  {124005} (\bibinfo {year} {2009})},\ \Eprint
  {https://arxiv.org/abs/0910.1279} {arXiv:0910.1279 [gr-qc]} \BibitemShut
  {NoStop}%
\bibitem [{\citenamefont {Eastwood}\ and\ \citenamefont
  {Singer}(1985)}]{EastwoodSinger}%
  \BibitemOpen
  \bibfield  {author} {\bibinfo {author} {\bibfnamefont {M.~G.}\ \bibnamefont
  {Eastwood}}\ and\ \bibinfo {author} {\bibfnamefont {M.}~\bibnamefont
  {Singer}},\ }\bibfield  {title} {\enquote {\bibinfo {title} {{A conformally
  invariant Maxwell gauge}},}\ }\href
  {https://doi.org/10.1016/0375-9601(85)90198-7} {\bibfield  {journal}
  {\bibinfo  {journal} {Phys. Lett.}\ }\textbf {\bibinfo {volume} {A107}},\
  \bibinfo {pages} {73--74} (\bibinfo {year} {1985})}\BibitemShut {NoStop}%
\bibitem [{\citenamefont {Paneitz}(1983)}]{Paneitz1983}%
  \BibitemOpen
  \bibfield  {author} {\bibinfo {author} {\bibfnamefont {S.}~\bibnamefont
  {Paneitz}},\ }\bibfield  {title} {\enquote {\bibinfo {title} {A quartic
  conformally covariant differential operator for arbitrary pseudo-riemannian
  manifolds},}\ }\href@noop {} {\bibfield  {journal} {\bibinfo  {journal}
  {Preprint}\ } (\bibinfo {year} {1983})}\BibitemShut {NoStop}%
\bibitem [{\citenamefont {Paneitz}(2008)}]{Paneitz2008}%
  \BibitemOpen
  \bibfield  {author} {\bibinfo {author} {\bibfnamefont {S.}~\bibnamefont
  {Paneitz}},\ }\bibfield  {title} {\enquote {\bibinfo {title} {A quartic
  conformally covariant differential operator for arbitrary pseudo-riemannian
  manifolds (summary)},}\ }\href {https://doi.org/10.3842/SIGMA.2008.036}
  {\bibfield  {journal} {\bibinfo  {journal} {Sigma}\ }\textbf {\bibinfo
  {volume} {4}},\ \bibinfo {pages} {036} (\bibinfo {year} {2008})},\ \Eprint
  {https://arxiv.org/abs/0803.4331} {Arxiv:0803.4331 [math.DG]} \BibitemShut
  {NoStop}%
\bibitem [{\citenamefont {Riegert}(1984)}]{Riegert1984}%
  \BibitemOpen
  \bibfield  {author} {\bibinfo {author} {\bibfnamefont {R.~J.}\ \bibnamefont
  {Riegert}},\ }\bibfield  {title} {\enquote {\bibinfo {title} {{A Nonlocal
  Action for the Trace Anomaly}},}\ }\href
  {https://doi.org/10.1016/0370-2693(84)90983-3} {\bibfield  {journal}
  {\bibinfo  {journal} {Phys. Lett.}\ }\textbf {\bibinfo {volume} {B134}},\
  \bibinfo {pages} {56--60} (\bibinfo {year} {1984})}\BibitemShut {NoStop}%
\bibitem [{\citenamefont {Fradkin}\ and\ \citenamefont
  {Tseytlin}(1984)}]{Fradkin1984}%
  \BibitemOpen
  \bibfield  {author} {\bibinfo {author} {\bibfnamefont {E.}~\bibnamefont
  {Fradkin}}\ and\ \bibinfo {author} {\bibfnamefont {A.~A.}\ \bibnamefont
  {Tseytlin}},\ }\bibfield  {title} {\enquote {\bibinfo {title} {{Conformal
  Anomaly in Weyl Theory and Anomaly Free Superconformal Theories}},}\ }\href
  {https://doi.org/10.1016/0370-2693(84)90668-3} {\bibfield  {journal}
  {\bibinfo  {journal} {Phys.Lett.}\ }\textbf {\bibinfo {volume} {B134}},\
  \bibinfo {pages} {187} (\bibinfo {year} {1984})}\BibitemShut {NoStop}%
\bibitem [{\citenamefont {Antoniadis}\ and\ \citenamefont
  {Mottola}(1992)}]{Antoniadis:1992}%
  \BibitemOpen
  \bibfield  {author} {\bibinfo {author} {\bibfnamefont {I.}~\bibnamefont
  {Antoniadis}}\ and\ \bibinfo {author} {\bibfnamefont {E.}~\bibnamefont
  {Mottola}},\ }\bibfield  {title} {\enquote {\bibinfo {title}
  {Four-dimensional quantum gravity in the conformal sector},}\ }\href
  {https://doi.org/10.1103/PhysRevD.45.2013} {\bibfield  {journal} {\bibinfo
  {journal} {Phys. Rev. D}\ }\textbf {\bibinfo {volume} {45}},\ \bibinfo
  {pages} {2013--2025} (\bibinfo {year} {1992})}\BibitemShut {NoStop}%
\bibitem [{\citenamefont {Bayen}(1970)}]{Bayen1970}%
  \BibitemOpen
  \bibfield  {author} {\bibinfo {author} {\bibfnamefont {F.}~\bibnamefont
  {Bayen}},\ }\href@noop {} {Ph.D. thesis},\ \bibinfo  {school} {Universit\'e
  de Dijon} (\bibinfo {year} {1970})\BibitemShut {NoStop}%
\bibitem [{\citenamefont {Drew}\ and\ \citenamefont
  {Gegenberg}(1980)}]{Drew:1980yk}%
  \BibitemOpen
  \bibfield  {author} {\bibinfo {author} {\bibfnamefont {M.~S.}\ \bibnamefont
  {Drew}}\ and\ \bibinfo {author} {\bibfnamefont {J.~D.}\ \bibnamefont
  {Gegenberg}},\ }\bibfield  {title} {\enquote {\bibinfo {title} {{Conformally
  covariant massless spin-2 field equations}},}\ }\href
  {https://doi.org/10.1007/BF02776555} {\bibfield  {journal} {\bibinfo
  {journal} {Nuovo Cim.}\ }\textbf {\bibinfo {volume} {A60}},\ \bibinfo {pages}
  {41--56} (\bibinfo {year} {1980})}\BibitemShut {NoStop}%
\bibitem [{\citenamefont {Barut}\ and\ \citenamefont
  {Xu}(1982)}]{Barut:1982nj}%
  \BibitemOpen
  \bibfield  {author} {\bibinfo {author} {\bibfnamefont {A.~O.}\ \bibnamefont
  {Barut}}\ and\ \bibinfo {author} {\bibfnamefont {B.-W.}\ \bibnamefont {Xu}},\
  }\bibfield  {title} {\enquote {\bibinfo {title} {{On Conformally Covariant
  Spin-2 and Spin-3/2 Equations}},}\ }\href
  {https://doi.org/10.1088/0305-4470/15/4/010} {\bibfield  {journal} {\bibinfo
  {journal} {J. Phys.}\ }\textbf {\bibinfo {volume} {A15}},\ \bibinfo {pages}
  {L207--L210} (\bibinfo {year} {1982})}\BibitemShut {NoStop}%
\bibitem [{\citenamefont {Leonovich}\ and\ \citenamefont
  {Nesterenko}(1984)}]{Leonovich:1984cf}%
  \BibitemOpen
  \bibfield  {author} {\bibinfo {author} {\bibfnamefont {A.~A.}\ \bibnamefont
  {Leonovich}}\ and\ \bibinfo {author} {\bibfnamefont {V.~V.}\ \bibnamefont
  {Nesterenko}},\ }\href@noop {} {\enquote {\bibinfo {title} {{Conformally
  invariant equation for the symmetric tensor}},}\ } (\bibinfo {year} {1984}),\
  \bibinfo {note} {jINR-E2-84-11}\BibitemShut {NoStop}%
\bibitem [{\citenamefont {Ben~Achour}, \citenamefont {Huguet},\ and\
  \citenamefont {Renaud}(2014)}]{Achour:2013afa}%
  \BibitemOpen
  \bibfield  {author} {\bibinfo {author} {\bibfnamefont {J.}~\bibnamefont
  {Ben~Achour}}, \bibinfo {author} {\bibfnamefont {E.}~\bibnamefont {Huguet}},\
  and\ \bibinfo {author} {\bibfnamefont {J.}~\bibnamefont {Renaud}},\
  }\bibfield  {title} {\enquote {\bibinfo {title} {{Conformally invariant wave
  equation for a symmetric second rank tensor (“spin-2”) in a
  $d$-dimensional curved background}},}\ }\href
  {https://doi.org/10.1103/PhysRevD.89.064041} {\bibfield  {journal} {\bibinfo
  {journal} {Phys. Rev.}\ }\textbf {\bibinfo {volume} {D89}},\ \bibinfo {pages}
  {064041} (\bibinfo {year} {2014})},\ \Eprint
  {https://arxiv.org/abs/1311.3124} {arXiv:1311.3124 [gr-qc]} \BibitemShut
  {NoStop}%
\bibitem [{\citenamefont {Delfino~Galles}(1985)}]{DelfinoGalles:1985bb}%
  \BibitemOpen
  \bibfield  {author} {\bibinfo {author} {\bibfnamefont {C.}~\bibnamefont
  {Delfino~Galles}},\ }\bibfield  {title} {\enquote {\bibinfo {title} {{A
  conformal gauge for the spin-2 field}},}\ }\href
  {https://doi.org/10.1007/BF02747059} {\bibfield  {journal} {\bibinfo
  {journal} {Lett. Nuovo Cim.}\ }\textbf {\bibinfo {volume} {42}},\ \bibinfo
  {pages} {382--384} (\bibinfo {year} {1985})}\BibitemShut {NoStop}%
\bibitem [{\citenamefont {W{\"u}nsch}(1986)}]{Wunsch86}%
  \BibitemOpen
  \bibfield  {author} {\bibinfo {author} {\bibfnamefont {V.}~\bibnamefont
  {W{\"u}nsch}},\ }\bibfield  {title} {\enquote {\bibinfo {title} {{On
  conformally invariant differential operators}},}\ }\href
  {https://doi.org/10.1002/mana.19861290123} {\bibfield  {journal} {\bibinfo
  {journal} {Math. Nachr.}\ }\textbf {\bibinfo {volume} {129}},\ \bibinfo
  {pages} {269–281} (\bibinfo {year} {1986})}\BibitemShut {NoStop}%
\bibitem [{\citenamefont {Brinkmann}(1925)}]{Brinkmann}%
  \BibitemOpen
  \bibfield  {author} {\bibinfo {author} {\bibfnamefont {H.~W.}\ \bibnamefont
  {Brinkmann}},\ }\bibfield  {title} {\enquote {\bibinfo {title} {{Einstein
  spaces which are mapped conformally on each other}},}\ }\href
  {https://doi.org/10.1007/BF01208647} {\bibfield  {journal} {\bibinfo
  {journal} {Math. Ann.}\ }\textbf {\bibinfo {volume} {94}},\ \bibinfo {pages}
  {119--145} (\bibinfo {year} {1925})}\BibitemShut {NoStop}%
\bibitem [{\citenamefont {Erdmenger}\ and\ \citenamefont
  {Osborn}(1998)}]{Erdmenger2}%
  \BibitemOpen
  \bibfield  {author} {\bibinfo {author} {\bibfnamefont {J.}~\bibnamefont
  {Erdmenger}}\ and\ \bibinfo {author} {\bibfnamefont {H.}~\bibnamefont
  {Osborn}},\ }\bibfield  {title} {\enquote {\bibinfo {title} {{Conformally
  covariant differential operators: Symmetric tensor fields}},}\ }\href
  {https://doi.org/10.1088/0264-9381/15/2/003} {\bibfield  {journal} {\bibinfo
  {journal} {Class. Quant. Grav.}\ }\textbf {\bibinfo {volume} {15}},\ \bibinfo
  {pages} {273--280} (\bibinfo {year} {1998})},\ \Eprint
  {https://arxiv.org/abs/gr-qc/9708040} {arXiv:gr-qc/9708040 [gr-qc]}
  \BibitemShut {NoStop}%
\bibitem [{\citenamefont {Wald}(1984)}]{WaldRG}%
  \BibitemOpen
  \bibfield  {author} {\bibinfo {author} {\bibfnamefont {R.~M.}\ \bibnamefont
  {Wald}},\ }\href {https://doi.org/10.7208/chicago/9780226870373.001.0001}
  {\emph {\bibinfo {title} {{General Relativity}}}}\ (\bibinfo  {publisher}
  {Chicago Univ. Pr.},\ \bibinfo {address} {Chicago, USA},\ \bibinfo {year}
  {1984})\BibitemShut {NoStop}%
\bibitem [{\citenamefont {Branson}(1995)}]{Branson1995}%
  \BibitemOpen
  \bibfield  {author} {\bibinfo {author} {\bibfnamefont {T.~P.}\ \bibnamefont
  {Branson}},\ }\bibfield  {title} {\enquote {\bibinfo {title} {{Sharp
  inequalities, the functional determinant, and the complementary series}},}\
  }\href {https://doi.org/10.1090/S0002-9947-1995-1316845-2} {\bibfield
  {journal} {\bibinfo  {journal} {Trans. Amer. Math. Soc.}\ }\textbf {\bibinfo
  {volume} {347}},\ \bibinfo {pages} {3671--3742} (\bibinfo {year}
  {1995})}\BibitemShut {NoStop}%
\bibitem [{\citenamefont {{Graham}}(2007)}]{Graham2007}%
  \BibitemOpen
  \bibfield  {author} {\bibinfo {author} {\bibfnamefont {C.~R.}\ \bibnamefont
  {{Graham}}},\ }\bibfield  {title} {\enquote {\bibinfo {title} {{Conformal
  Powers of the Laplacian via Stereographic Projection}},}\ }\href
  {https://doi.org/10.3842/SIGMA.2007.121} {\bibfield  {journal} {\bibinfo
  {journal} {SIGMA}\ }\textbf {\bibinfo {volume} {3}},\ \bibinfo {pages} {121}
  (\bibinfo {year} {2007})},\ \Eprint {https://arxiv.org/abs/0711.4798}
  {arXiv:0711.4798 [math.DG]} \BibitemShut {NoStop}%
\bibitem [{\citenamefont {Graham}\ \emph {et~al.}(1992)\citenamefont {Graham},
  \citenamefont {Jenne}, \citenamefont {Mason},\ and\ \citenamefont
  {Sparling}}]{GJMS1}%
  \BibitemOpen
  \bibfield  {author} {\bibinfo {author} {\bibfnamefont {C.}~\bibnamefont
  {Graham}}, \bibinfo {author} {\bibfnamefont {R.}~\bibnamefont {Jenne}},
  \bibinfo {author} {\bibfnamefont {L.}~\bibnamefont {Mason}},\ and\ \bibinfo
  {author} {\bibfnamefont {G.}~\bibnamefont {Sparling}},\ }\bibfield  {title}
  {\enquote {\bibinfo {title} {{Conformally Invariant Powers of the Laplacian,
  I: {E}xistence}},}\ }\href {https://doi.org/10.1112/jlms/s2-46.3.557}
  {\bibfield  {journal} {\bibinfo  {journal} {J. London Math. Soc.}\ }\textbf
  {\bibinfo {volume} {46}},\ \bibinfo {pages} {557--565} (\bibinfo {year}
  {1992})}\BibitemShut {NoStop}%
\bibitem [{\citenamefont {Graham}(1992)}]{GJMS2}%
  \BibitemOpen
  \bibfield  {author} {\bibinfo {author} {\bibfnamefont {C.}~\bibnamefont
  {Graham}},\ }\bibfield  {title} {\enquote {\bibinfo {title} {{Conformally
  Invariant Powers of the Laplacian, II: {N}onexistence}},}\ }\href
  {https://doi.org/10.1112/jlms/s2-46.3.566} {\bibfield  {journal} {\bibinfo
  {journal} {J. London Math. Soc.}\ }\textbf {\bibinfo {volume} {46}},\
  \bibinfo {pages} {566--576} (\bibinfo {year} {1992})}\BibitemShut {NoStop}%
\bibitem [{\citenamefont {Gover}\ and\ \citenamefont {Hirachi}(2004)}]{GJMS3}%
  \BibitemOpen
  \bibfield  {author} {\bibinfo {author} {\bibfnamefont {A.~R.}\ \bibnamefont
  {Gover}}\ and\ \bibinfo {author} {\bibfnamefont {K.}~\bibnamefont
  {Hirachi}},\ }\bibfield  {title} {\enquote {\bibinfo {title} {{Conformally
  invariant powers of the Laplacian: A Complete non-existence theorem}},}\
  }\href {https://doi.org/10.1090/S0894-0347-04-00450-3} {\bibfield  {journal}
  {\bibinfo  {journal} {J. Am. Math. Soc.}\ }\textbf {\bibinfo {volume} {17}},\
  \bibinfo {pages} {389--405} (\bibinfo {year} {2004})},\ \Eprint
  {https://arxiv.org/abs/math/0304082} {arXiv:math/0304082 [math-dg]}
  \BibitemShut {NoStop}%
\bibitem [{\citenamefont {Jakobsen}\ and\ \citenamefont
  {Vergne}(1977)}]{JakobsenVergne}%
  \BibitemOpen
  \bibfield  {author} {\bibinfo {author} {\bibfnamefont {H.}~\bibnamefont
  {Jakobsen}}\ and\ \bibinfo {author} {\bibfnamefont {M.}~\bibnamefont
  {Vergne}},\ }\bibfield  {title} {\enquote {\bibinfo {title} {{Wave and Dirac
  Operators, and Representations of the Conformal Group}},}\ }\href
  {https://doi.org/10.1016/0022-1236(77)90005-2} {\bibfield  {journal}
  {\bibinfo  {journal} {J. Funct. Anal.}\ }\textbf {\bibinfo {volume} {24}},\
  \bibinfo {pages} {52--106} (\bibinfo {year} {1977})}\BibitemShut {NoStop}%
\bibitem [{\citenamefont {Huguet}, \citenamefont {Queva},\ and\ \citenamefont
  {Renaud}(2006)}]{Huguet:2006fe}%
  \BibitemOpen
  \bibfield  {author} {\bibinfo {author} {\bibfnamefont {E.}~\bibnamefont
  {Huguet}}, \bibinfo {author} {\bibfnamefont {J.}~\bibnamefont {Queva}},\ and\
  \bibinfo {author} {\bibfnamefont {J.}~\bibnamefont {Renaud}},\ }\bibfield
  {title} {\enquote {\bibinfo {title} {{Conformally related massless fields in
  dS, AdS and Minkowski spaces}},}\ }\href
  {https://doi.org/10.1103/PhysRevD.73.084025} {\bibfield  {journal} {\bibinfo
  {journal} {Phys. Rev.}\ }\textbf {\bibinfo {volume} {D73}},\ \bibinfo {pages}
  {084025} (\bibinfo {year} {2006})},\ \Eprint
  {https://arxiv.org/abs/gr-qc/0603031} {arXiv:gr-qc/0603031 [gr-qc]}
  \BibitemShut {NoStop}%
\bibitem [{\citenamefont {Dixmier}(1961)}]{Dixmier}%
  \BibitemOpen
  \bibfield  {author} {\bibinfo {author} {\bibfnamefont {J.}~\bibnamefont
  {Dixmier}},\ }\bibfield  {title} {\enquote {\bibinfo {title}
  {Repr\'esentations int\'egrables du groupe de {De} {Sitter}},}\ }\href
  {https://doi.org/10.24033/bsmf.1558} {\bibfield  {journal} {\bibinfo
  {journal} {Bull. Soc. Math. France}\ }\textbf {\bibinfo {volume} {89}},\
  \bibinfo {pages} {9--41} (\bibinfo {year} {1961})}\BibitemShut {NoStop}%
\bibitem [{\citenamefont {Takahashi}(1963)}]{Takahashi}%
  \BibitemOpen
  \bibfield  {author} {\bibinfo {author} {\bibfnamefont {R.}~\bibnamefont
  {Takahashi}},\ }\bibfield  {title} {\enquote {\bibinfo {title} {Sur les
  repr\'esentations unitaires des groupes de {Lorentz} g\'en\'eralis\'es},}\
  }\href {https://doi.org/10.24033/bsmf.1598} {\bibfield  {journal} {\bibinfo
  {journal} {Bull. Soc. Math. France}\ }\textbf {\bibinfo {volume} {91}},\
  \bibinfo {pages} {289--433} (\bibinfo {year} {1963})}\BibitemShut {NoStop}%
\bibitem [{\citenamefont {Allen}(1985)}]{Allen1985}%
  \BibitemOpen
  \bibfield  {author} {\bibinfo {author} {\bibfnamefont {B.}~\bibnamefont
  {Allen}},\ }\bibfield  {title} {\enquote {\bibinfo {title} {{Vacuum States in
  de Sitter Space}},}\ }\href {https://doi.org/10.1103/PhysRevD.32.3136}
  {\bibfield  {journal} {\bibinfo  {journal} {Phys. Rev.}\ }\textbf {\bibinfo
  {volume} {D32}},\ \bibinfo {pages} {3136} (\bibinfo {year}
  {1985})}\BibitemShut {NoStop}%
\bibitem [{\citenamefont {Allen}\ and\ \citenamefont
  {Folacci}(1987)}]{Allen1987}%
  \BibitemOpen
  \bibfield  {author} {\bibinfo {author} {\bibfnamefont {B.}~\bibnamefont
  {Allen}}\ and\ \bibinfo {author} {\bibfnamefont {A.}~\bibnamefont
  {Folacci}},\ }\bibfield  {title} {\enquote {\bibinfo {title} {{The Massless
  Minimally Coupled Scalar Field in De Sitter Space}},}\ }\href
  {https://doi.org/10.1103/PhysRevD.35.3771} {\bibfield  {journal} {\bibinfo
  {journal} {Phys. Rev.}\ }\textbf {\bibinfo {volume} {D35}},\ \bibinfo {pages}
  {3771} (\bibinfo {year} {1987})}\BibitemShut {NoStop}%
\bibitem [{\citenamefont {Kirsten}\ and\ \citenamefont
  {Garriga}(1993)}]{Kirsten:1993ug}%
  \BibitemOpen
  \bibfield  {author} {\bibinfo {author} {\bibfnamefont {K.}~\bibnamefont
  {Kirsten}}\ and\ \bibinfo {author} {\bibfnamefont {J.}~\bibnamefont
  {Garriga}},\ }\bibfield  {title} {\enquote {\bibinfo {title} {{Massless
  minimally coupled fields in de Sitter space: O(4) symmetric states versus de
  Sitter invariant vacuum}},}\ }\href {https://doi.org/10.1103/PhysRevD.48.567}
  {\bibfield  {journal} {\bibinfo  {journal} {Phys. Rev.}\ }\textbf {\bibinfo
  {volume} {D48}},\ \bibinfo {pages} {567--577} (\bibinfo {year} {1993})},\
  \Eprint {https://arxiv.org/abs/gr-qc/9305013} {arXiv:gr-qc/9305013 [gr-qc]}
  \BibitemShut {NoStop}%
\bibitem [{\citenamefont {Gazeau}, \citenamefont {Renaud},\ and\ \citenamefont
  {Takook}(2000)}]{Gazeau:1999mi}%
  \BibitemOpen
  \bibfield  {author} {\bibinfo {author} {\bibfnamefont {J.-P.}\ \bibnamefont
  {Gazeau}}, \bibinfo {author} {\bibfnamefont {J.}~\bibnamefont {Renaud}},\
  and\ \bibinfo {author} {\bibfnamefont {M.~V.}\ \bibnamefont {Takook}},\
  }\bibfield  {title} {\enquote {\bibinfo {title} {{Gupta-Bleuler quantization
  for minimally coupled scalar fields in de Sitter space}},}\ }\href
  {https://doi.org/10.1088/0264-9381/17/6/307} {\bibfield  {journal} {\bibinfo
  {journal} {Class. Quant. Grav.}\ }\textbf {\bibinfo {volume} {17}},\ \bibinfo
  {pages} {1415--1434} (\bibinfo {year} {2000})},\ \Eprint
  {https://arxiv.org/abs/gr-qc/9904023} {arXiv:gr-qc/9904023 [gr-qc]}
  \BibitemShut {NoStop}%
\bibitem [{\citenamefont {Joung}, \citenamefont {Mourad},\ and\ \citenamefont
  {Parentani}(2007)}]{Joung:2007je}%
  \BibitemOpen
  \bibfield  {author} {\bibinfo {author} {\bibfnamefont {E.}~\bibnamefont
  {Joung}}, \bibinfo {author} {\bibfnamefont {J.}~\bibnamefont {Mourad}},\ and\
  \bibinfo {author} {\bibfnamefont {R.}~\bibnamefont {Parentani}},\ }\bibfield
  {title} {\enquote {\bibinfo {title} {{Group theoretical approach to quantum
  fields in de Sitter space. II. The complementary and discrete series}},}\
  }\href {https://doi.org/10.1088/1126-6708/2007/09/030} {\bibfield  {journal}
  {\bibinfo  {journal} {JHEP}\ }\textbf {\bibinfo {volume} {09}},\ \bibinfo
  {pages} {030} (\bibinfo {year} {2007})},\ \Eprint
  {https://arxiv.org/abs/0707.2907} {arXiv:0707.2907 [hep-th]} \BibitemShut
  {NoStop}%
\bibitem [{\citenamefont {Folacci}(1992)}]{Folacci:1992xc}%
  \BibitemOpen
  \bibfield  {author} {\bibinfo {author} {\bibfnamefont {A.}~\bibnamefont
  {Folacci}},\ }\bibfield  {title} {\enquote {\bibinfo {title} {{BRST
  quantization of the massless minimally coupled scalar field in de Sitter
  space: Zero modes, euclideanization and quantization}},}\ }\href
  {https://doi.org/10.1103/PhysRevD.46.2553} {\bibfield  {journal} {\bibinfo
  {journal} {Phys. Rev.}\ }\textbf {\bibinfo {volume} {D46}},\ \bibinfo {pages}
  {2553--2559} (\bibinfo {year} {1992})},\ \Eprint
  {https://arxiv.org/abs/0911.2064} {arXiv:0911.2064 [gr-qc]} \BibitemShut
  {NoStop}%
\bibitem [{\citenamefont {Folacci}(1996)}]{Folacci:1996dv}%
  \BibitemOpen
  \bibfield  {author} {\bibinfo {author} {\bibfnamefont {A.}~\bibnamefont
  {Folacci}},\ }\bibfield  {title} {\enquote {\bibinfo {title} {{Toy model for
  the zero mode problem in the conformal sector of de Sitter quantum
  gravity}},}\ }\href {https://doi.org/10.1103/PhysRevD.53.3108} {\bibfield
  {journal} {\bibinfo  {journal} {Phys. Rev.}\ }\textbf {\bibinfo {volume}
  {D53}},\ \bibinfo {pages} {3108--3117} (\bibinfo {year} {1996})}\BibitemShut
  {NoStop}%
\bibitem [{\citenamefont {Bros}, \citenamefont {Epstein},\ and\ \citenamefont
  {Moschella}(2010)}]{Bros:2010wa}%
  \BibitemOpen
  \bibfield  {author} {\bibinfo {author} {\bibfnamefont {J.}~\bibnamefont
  {Bros}}, \bibinfo {author} {\bibfnamefont {H.}~\bibnamefont {Epstein}},\ and\
  \bibinfo {author} {\bibfnamefont {U.}~\bibnamefont {Moschella}},\ }\bibfield
  {title} {\enquote {\bibinfo {title} {{Scalar tachyons in the de Sitter
  universe}},}\ }\href {https://doi.org/10.1007/s11005-010-0406-4} {\bibfield
  {journal} {\bibinfo  {journal} {Lett. Math. Phys.}\ }\textbf {\bibinfo
  {volume} {93}},\ \bibinfo {pages} {203--211} (\bibinfo {year} {2010})},\
  \Eprint {https://arxiv.org/abs/1003.1396} {arXiv:1003.1396 [hep-th]}
  \BibitemShut {NoStop}%
\end{thebibliography}%
\end{document}